\documentclass[authoryear]{elsarticle}

\usepackage{hyperref}
\usepackage{bm}
\usepackage{amsmath}
\usepackage{amssymb}
\usepackage{subfigure}
\usepackage{color}
\usepackage{graphicx}

\journal{Icarus}

\begin{document}

\begin{frontmatter}

  \title{Collisional Elongation: Possible Origin of Extremely Elongated Shape of 1I/`Oumuamua}

\author[mymainaddress]{Keisuke Sugiura\corref{mycorrespondingauthor}}
\cortext[mycorrespondingauthor]{Corresponding author}
\ead{sugiura.keisuke@a.mbox.nagoya-u.ac.jp}

\author[mymainaddress]{Hiroshi Kobayashi}

\author[mymainaddress]{Shu-ichiro Inutsuka}

\address[mymainaddress]{Department of Physics, Nagoya University, Aichi 464-8602, Japan}

\begin{abstract}
  Light curve observations of a recently discovered interstellar object 1I/`Oumuamua suggest that this object has an extremely elongated shape with the axis ratio 0.3 or smaller. Planetesimal collisions can produce irregular shapes including elongated shapes. In this paper, we suggest that the extremely elongated shape of 1I/`Oumuamua may be the result of such an impact. To find detailed impact conditions to form the extremely elongated objects, we conduct numerical simulations of planetesimal collisions using Smoothed Particle Hydrodynamics method for elastic dynamics with self-gravity and interparticle friction. Impacts into strengthless target planetesimals with radius $50\,{\rm m}$ are conducted with various ratios of impactor mass to target mass $q$, friction angles $\phi_{{\rm d}}$, impact velocities $v_{{\rm imp}}$, and impact angles $\theta_{{\rm imp}}$. We find that impacts with $q \geq 0.5$, $\phi_{{\rm d}} \geq 40^{\circ}$, $v_{{\rm imp}} \leq 40\,{\rm cm/s}$, and $\theta_{{\rm imp}} \leq 30^{\circ}$ produce remnants with the ratio of intermediate to major axis length less than 0.3. This impact condition suggests that the parent protoplanetary disk in the planetesimal collision stage was weakly turbulent ($\alpha < 10^{-4}$ for the inner disk) and composed of planetesimals smaller than $\sim 7\,{\rm km}$ to ensure small impact velocity.
\end{abstract}


\end{frontmatter}

\section{Introduction}

1I/`Oumuamua was discovered by Pan-STARRS1 on October 2017 (\citealt{Meech-et-al2017}). It is recognized to be an interstellar object because of its high orbital eccentricity $e \approx 1.2$. A considerable fraction of planetesimals are ejected from planetary systems during the planet formation stage (\citealt{Fernandez1978, Bottke-et-al2005, Raymond-et-al2017, Jackson-et-al2017}); thus it is no wonder that interstellar objects exist. Moreover, the velocity at infinity $v_{\infty} \approx 26\,{\rm km/s}$ of 1I/`Oumuamua is close to relative velocities of stars in the solar neighborhood. Therefore this object is likely to come from a nearby planetary system (\citealt{Meech-et-al2017, de-la-Fuente-Marcos-double2017}). 

The absolute magnitude of 1I/`Oumuamua is $H_{{\rm V}} \approx 22\,{\rm mag}$, which corresponds to a mean radius of about $100\,{\rm m}$ with albedo 0.04 (\citealt{Meech-et-al2017, Bolin-et-al2017, Bannister-et-al2017}). The rotation period of 1I/`Oumuamua is about 8 hours (e.g., \citealt{Bolin-et-al2017}). Its spectral features are consistent with that of D-type asteroids or comets (\citealt{Meech-et-al2017}). Although no cometary activity is reported based on the analyses of imaging observations (\citealt{Meech-et-al2017, Ye-et-al2017, Jewitt-et-al2017}), non-gravitational acceleration of 1I/`Oumuamua is measured from its trajectory (\citealt{Micheli-et-al2018}), which may be the result of mild cometary activity. Thus it is not clear whether 1I/`Oumuamua is a volatile-rich comet-like body or a rocky asteroid-like body.

An interesting characteristic of 1I/`Oumuamua is the large light curve amplitude of about $2.5\,{\rm mag}$ (\citealt{Meech-et-al2017}). If this light curve variation is caused by change of cross section due to rotation around the minor axis, 1I/`Oumuamua has an extremely elongated shape. Although its dimensions are not yet fully specified because of the short observational period, the ratio of intermediate axis length $b$ to major axis length $a$ is estimated in several studies as follows: $b/a < 0.19$ (\citealt{Bannister-et-al2017}), $0.10 < b/a < 0.29$ (\citealt{Bolin-et-al2017}), $b/a < 0.22$ (\citealt{Drahus-et-al2018}), $b/a < 0.20$ (\citealt{Fraser-et-al2017}), $b/a < 0.33$ (\citealt{Knight-et-al2017}), and $b/a \approx 0.1$ (\citealt{Meech-et-al2017}). Therefore, the ratio $b/a$ is estimated to be smaller than about 0.3.

Several ideas have been proposed in the past to produce/explain extremely elongated objects. For example, impact experiments in laboratories show that some fragments among several thousands may have $b/a < 0.3$ (\citealt{Michikami-et-al2016}). Tidal disruption may also produce elongated objects (\citealt{Cuk2017}). Simulations for the tidal destruction of $3\,{\rm km}$ sized rubble piles reproduced with $N$-body particles show that about 0.5\% of resultant bodies produced in all simulations have $b/a<0.3$ (\citealt{Walsh-and-Richardson2006}). Simulations for the spin deformation of rubble piles show that rubble piles with spin states close to equilibrium limits may evolve to elongated shapes (\citealt{Richardson-et-al2005}). Recently it is suggested that extremely elongated objects may be formed through abrasion due to micro particle collisions (\citealt{Domokos-et-al2009, Domokos-et-al2017}). Finally, collisional fragmentation and gravitational accumulation can form elongated objects (e.g.,\,\citealt{Leinhardt-et-al2000,Leinhardt-et-al2010,Sugiura-et-al2018}).

Collisions generally occur during plant formation. In this paper, we investigate this collisional process in significant detail, namely, the possibility of forming extremely elongated objects through collisions. In our previous work (\citealt{Sugiura-et-al2018}), we conducted numerical simulations of equal-mass impacts between $50\,{\rm km}$ radius planetesimals, investigated the impact conditions required to form various irregular shapes, and showed that objects with $b/a \approx 0.2$ can be formed through planetesimal collisions. Here, we focus on impact simulations with much smaller ($50\,{\rm m}$ sized) planetesimals to reproduce extremely elongated objects with the size similar to that of 1I/`Oumuamua, and constrain detailed impact conditions to form shapes with $b/a < 0.3$. Here, we define remnants with $b/a<0.3$ as extremely elongated remnants (hereafter EERs). 

\section{Method and initial conditions}
To investigate the formation of irregularly shaped objects through planetesimal collisions and gravitational reaccumulation of fragments, we use a Smoothed Particle Hydrodynamics (SPH) code for elastic dynamics (\citealt{Libersky-and-Petschek1991}) that includes self-gravity, a fracture model for rocky material (Grady-Kipp fragmentation model: \citealt{Benz-and-Asphaug1995}), and a friction model for granular material (Drucker-Prager yield criterion with zero cohesion: \citealt{Jutzi2015}). We parallelize our code utilizing Framework for Developing Particle Simulator (\citealt{Iwasawa-et-al2015,Iwasawa-et-al2016}). We use basaltic spheres with zero rotation as colliding planetesimals. We conduct each impact simulation until $1.0\times 10^{5}\,{\rm s}$ after the time of impact, which is about 100 times longer than the typical timescale of reaccumulation $\sim 10^{3}{\rm s}$. Thus the deformation of shapes due to reaccumulation of fragments is finished in much less than $1.0\times 10^{5}\,{\rm s}$. After each impact simulation, we measure axis ratios of remnants based on the procedure described in \cite{Sugiura-et-al2018}.

In the rest of this section, we describe the highlight of the numerical methods, collision parameters, and initial conditions that differ from our previous work \cite{Sugiura-et-al2018}. For a more detailed description of our methods, see \cite{Sugiura-et-al2018}.

For a kernel function $W(|\bm{x}_{i}-\bm{x}_{j}|,h)$, we use a cubic spline kernel (e.g.,\,\citealt{Monaghan-and-Lattanzio1985}) given by 

\begin{eqnarray}
  W(|\bm{x}_{i}-\bm{x}_{j}|,h)=\frac{1}{\pi h^{3}}\left\{ \begin{array}{ll}
    1-\frac{3}{2}\Bigl(\frac{|\bm{x}_{i}-\bm{x}_{j}|}{h}\Bigr)^{2}+\frac{3}{4}\Bigl(\frac{|\bm{x}_{i}-\bm{x}_{j}|}{h}\Bigr)^{3} & 0 \leq \frac{|\bm{x}_{i}-\bm{x}_{j}|}{h} < 1 \\
    \frac{1}{4}\Bigl(2-\frac{|\bm{x}_{i}-\bm{x}_{j}|}{h}\Bigr)^{3} & 1 \leq \frac{|\bm{x}_{i}-\bm{x}_{j}|}{h} < 2 \\
    0 & 2 \leq \frac{|\bm{x}_{i}-\bm{x}_{j}|}{h}  \\
  \end{array} \right. ,
  \label{kernel-function}
\end{eqnarray}

\noindent where $h$ is the smoothing length and $\bm{x}_{i}$ and $\bm{x}_{j}$ are the position vectors of the $i$-th and $j$-th SPH particles, respectively. The kernel has a support radius smaller than that for Gaussian kernel, which reduces computational cost and enables investigation of a wider parameter space.

For collisional formation of EERs with sizes similar to 1I/`Oumuamua, the radius of target planetesimals is set to $R_{{\rm t}}=50\,{\rm m}$. We do not limit our simulations to equal-mass impacts, and simulations with several mass ratios $q=M_{{\rm i}}/M_{{\rm t}}$ are also performed, where $M_{{\rm t}}$ is the mass of targets, and $M_{{\rm i}} (\leq M_{{\rm t}})$ is the mass of impactors. We use about 50,000 SPH particles for a target planetesimal because \cite{Sugiura-et-al2018} showed that this number of particles is sufficient to achieve the converged value of axis ratios of the largest remnants through the simulations of merging collisions with various resolutions.

We assume that the initial impactors are small primordial planetesimals that are too small to have melted due to radioactive decay of $^{26}{\rm Al}$ (\citealt{Wakita-et-al2014}). As a result, our initial planetesimals are set to have no tensile strength of monolith body but they do have shear strength determined by the friction of granular material (\citealt{Richardson-et-al2002}). Following the fracture model of \cite{Benz-and-Asphaug1995}, we use a damage parameter $D$. An SPH particle with $D=0$ represents intact rock, and that with $D=1$ represents completely disrupted rock, or granular material. The damage parameter $D$ of all SPH particles is initially set to unity.

Frictional force depends on the friction coefficient $\mu_{{\rm d}}=\tan (\phi_{{\rm d}})$, where $\phi_{{\rm d}}$ is the friction angle, which is generally the same as the angle of repose. The friction angle of lunar soil is estimated to be $30^{\circ} - 50^{\circ}$ (\citealt{Heiken-et-al1991}). We vary $\phi_{{\rm d}}$ in this range.

We use the Tillotson equation of state (EOS) for basalt described in \cite{Benz-and-Asphaug1999}. For low internal energy in the Tillotson EOS, the pressure $p$ is given by

\begin{equation}
  p=\Bigl[ a_{{\rm Til}}+\frac{b_{{\rm Til}}}{(u\rho_{0}^{2}/u_{0,{\rm Til}}\rho^{2})+1} \Bigr]\rho u +A_{{\rm Til}}\Bigl( \frac{\rho}{\rho_{0}}-1 \Bigr) + B_{{\rm Til}} \Bigl( \frac{\rho}{\rho_{0}}-1 \Bigr)^{2},
  \label{Tillotson-eos}
\end{equation}

\noindent where $\rho_{0}$ is the uncompressed density, $a_{{\rm Til}}$, $b_{{\rm Til}}$, $A_{{\rm Til}}$, $B_{{\rm Til}}$, and $u_{0,{\rm Til}}$ are the material-dependent Tillotson parameters, and $\rho$ and $u$ are the density and specific internal energy, respectively. Note that $A_{{\rm Til}} \simeq B_{{\rm Til}}$ for most geologic materials. For impacts considered here, $\rho \sim \rho_{0}$ and $u \sim 0$, and hence the sound speed $C_{{\rm s}}$ is given by $\sqrt{A_{{\rm Til}}/\rho_{0}}$. 

A time step determined by the sound speed of basaltic body $C_{{\rm s}} \approx 3\,{\rm km/s}$ is calculated to be $\sim 10^{-3}\,{\rm s}$, which is much smaller than the timescale of gravitational reaccumulation estimated to be $2R_{{\rm t}}/v_{{\rm esc}}$ $\approx 1,600{\rm s}$, where $v_{{\rm esc}}$ is two-body escape velocity of impacting planetesimals. Thus it is difficult to complete a simulation in an acceptable computation time with $C_{{\rm s}} \approx 3\,{\rm km/s}$. However, if impact velocities are much smaller than $C_{{\rm s}}$, the shapes of collisional remnants obtained from simulations are almost independent of the value of $C_{{\rm s}}$ because the shear strength is mainly determined by the friction (see Appendix A). Thus we set $A_{{\rm Til}}$ and $B_{{\rm Til}}$ to $2.67\times 10^{4}\,{\rm Pa}$, which corresponds to $C_{{\rm s}}\approx 3\,{\rm m/s}$. This sound speed is still much larger than impact velocities treated in our simulations $\leq 40\,{\rm cm/s}$. The same approach utilizing reduced sound speed is adopted in \cite{Jutzi-and-Asphaug2015}. The values of the Tillotson parameters used in the simulations are listed in Table \ref{table-Tillotson-parameters}.

\begin{table}[!htb]
  \begin{center}
    \begin{tabular}{c c c c c c}\hline 
      $\rho_{0}{\rm (kg/m^{3})}$ & $A_{{\rm Til}}{\rm (Pa)}$ & $B_{{\rm Til}}{\rm (Pa)}$ & $u_{0,{\rm Til}}{\rm (J/kg)}$ & $a_{{\rm Til}}$ & $b_{{\rm Til}}$ \\
      \hline 
      $2.7\times 10^{3}$  & $2.67\times 10^{4}$ & $2.67\times 10^{4}$ & $4.87\times 10^{8}$ & $0.5$ & $1.5$ \\ 
      \hline
    \end{tabular}
  \end{center}
  \caption{Tillotson parameters used in the simulations.}
  \label{table-Tillotson-parameters}
\end{table}

\section{Results}
We conduct simulations with various mass ratios $q$, impact angles $\theta_{{\rm imp}}$, impact velocities $v_{{\rm imp}}$, and friction angles $\phi_{{\rm d}}$. Table \ref{simulation-conditinos} summarizes the conditions of the simulations.

\begin{table}[!htb]
  \begin{center}
    \begin{tabular}{c c c c c c c}\hline 
      $q$ & $\phi_{{\rm d}}(^{\circ})$ & $v_{{\rm imp}}({\rm cm/s})$ & $\Delta v_{{\rm imp}}({\rm cm/s})$ & $\theta_{{\rm imp}}(^{\circ})$ & $\Delta \theta_{{\rm imp}}(^{\circ})$ & Remark \\
      \hline 
      $1$  & $40$ & $12 - 36$ & $3$ & $5 - 30$ & $5$ & Fig.\ref{b-a-of-largest-remnant-q=1.0-phid=40}a \\
      $1$  & $40$ & $15 - 30$ & $1$ & $7.5 - 20$ & $2.5$ & Fig.\ref{b-a-of-largest-remnant-q=1.0-phid=40}b \\
      $1$  & $50$ & $12.5 - 40$ & $2.5$ & $5 - 40$ & $5$ & Fig.\ref{b-a-of-largest-remnant-q=1.0-phid=50} \\
      $1$  & $30$ & $12.5 - 35$ & $2.5$ & $5 - 20$ & $5$ &  \\
      $0.5$  & $50$ & $15 - 35$ & $2$ & $10 - 35$ & $5$ & Fig.\ref{b-a-of-largest-remnant-q=0.5-phid=50} \\
      $0.25$  & $50$ & $15 - 55$ & $4$ & $10 - 35$ & $5$ &  \\
      \hline
    \end{tabular}
  \end{center}
  \caption{Impact conditions of the simulations. $\Delta v_{{\rm imp}}$ and $\Delta \theta_{{\rm imp}}$ are the increments of $v_{{\rm imp}}$ and $\theta_{{\rm imp}}$ in the parameter survey, respectively.}
  \label{simulation-conditinos}
\end{table}

\subsection{Equal-mass impacts with $\phi_{{\rm d}}=40^{\circ}$}

\begin{figure}[!htb]
  \begin{center}
    \includegraphics[bb=0 0 960 498, width=1.0\linewidth,clip]{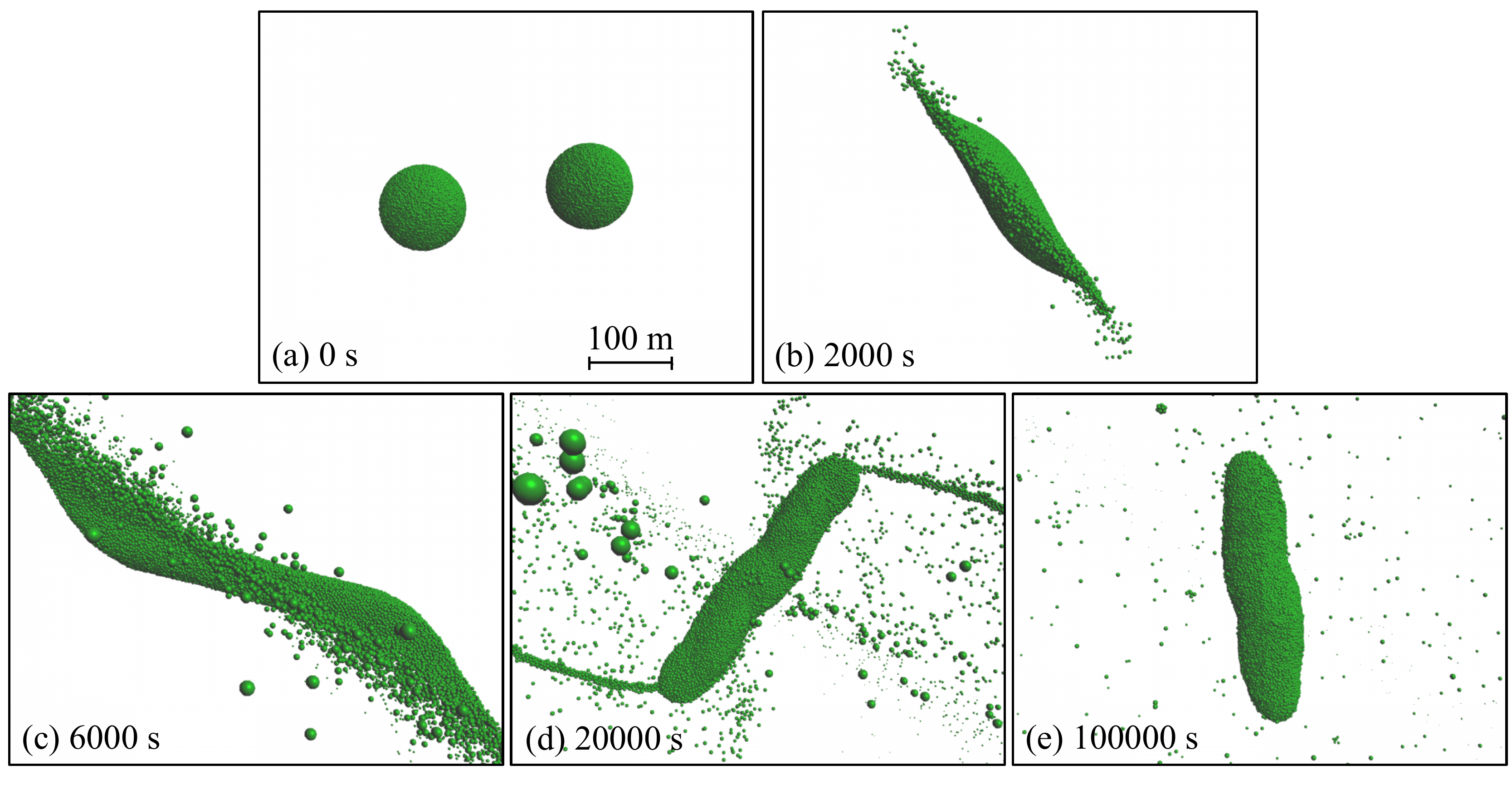}
    \caption{Snapshots of the impact simulation with $q=1$, $\theta_{{\rm imp}}=15^{\circ}$, $v_{{\rm imp}}=20\,{\rm cm/s}$, and $\phi_{{\rm d}}=40^{\circ}$. }
    \label{subsequent-pictures-50m-1.0-15-20cms-phid=40}
  \end{center}
\end{figure}

Figure \ref{subsequent-pictures-50m-1.0-15-20cms-phid=40} shows snapshots of the impact simulation with $q=1$, $\theta_{{\rm imp}}=15^{\circ}$, $v_{{\rm imp}}=20\,{\rm cm/s}$, and $\phi_{{\rm d}}=40^{\circ}$. The collision induces elongation of the body (Fig.\ref{subsequent-pictures-50m-1.0-15-20cms-phid=40}b,c), which leads to the formation of an EER (Fig.\ref{subsequent-pictures-50m-1.0-15-20cms-phid=40}d,e). The ratio $b/a$ of the largest remnant is 0.24. The rotation period of the largest remnant is 9.36 hours, which is comparable to that of 1I/`Oumuamua. As shown in Fig. \ref{subsequent-pictures-50m-1.0-15-20cms-phid=40}, we confirm that the EER keeps its extremely elongated shape at least until $1.0\times 10^{5}$ s that corresponds to about three rotations of this object.

\begin{figure}[!htb]
  \begin{center}
    \includegraphics[bb=0 0 808 1077, width=0.6\linewidth,clip]{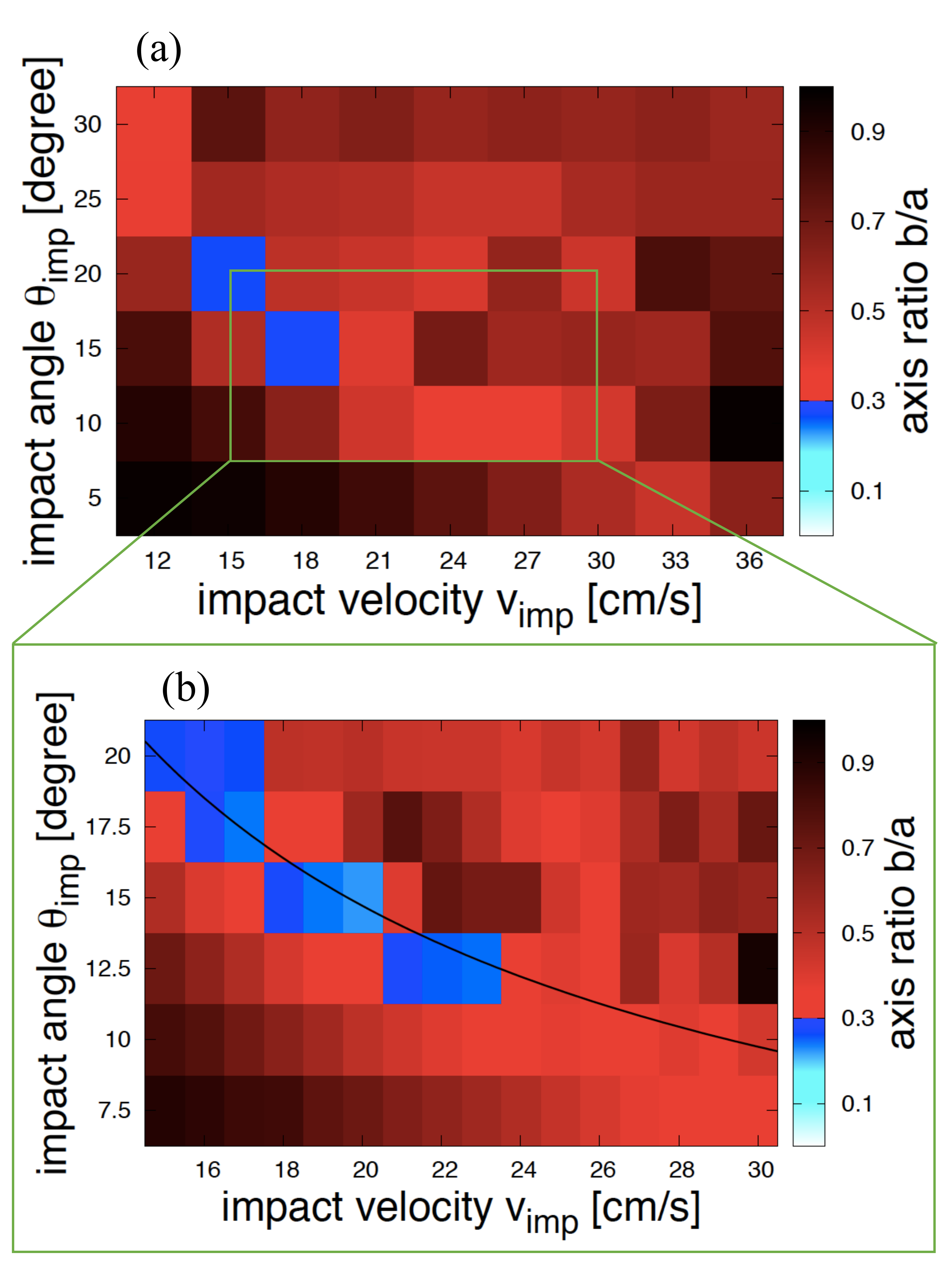}
    \caption{$b/a$ of the largest remnants resulting from various impact simulations with $q=1$ and $\phi_{{\rm d}}=40^{\circ}$. Horizontal axis shows $v_{{\rm imp}}$, and vertical axis shows $\theta_{{\rm imp}}$. Blue regions show parameters forming EERs, which have $b/a<0.3$. In (a), we vary parameters within $12\,{\rm cm/s} \leq v_{{\rm imp}} \leq 36\,{\rm cm/s}$ and $5^{\circ} \leq \theta_{{\rm imp}} \leq 30^{\circ}$ with increments $3\,{\rm cm/s}$ and $5^{\circ}$. In (b), we vary parameters within $15\,{\rm cm/s} \leq v_{{\rm imp}} \leq 30\,{\rm cm/s}$ and $7.5^{\circ} \leq \theta_{{\rm imp}} \leq 20^{\circ}$ with increments $1\,{\rm cm/s}$ and $2.5^{\circ}$. Black curve in (b) represents $v_{{\rm imp}}\sin \theta_{{\rm imp}} = 5.1\,{\rm cm/s}$.}
    \label{b-a-of-largest-remnant-q=1.0-phid=40}
  \end{center}
\end{figure}

Figure \ref{b-a-of-largest-remnant-q=1.0-phid=40} shows $b/a$ of the largest remnants, which mainly become the most elongated objects, formed through impacts with $q=1$ and $\phi_{{\rm d}}=40^{\circ}$. From Fig.\,\ref{b-a-of-largest-remnant-q=1.0-phid=40}a, we find that EERs are only formed for $\theta_{{\rm imp}}\approx 15^{\circ}$ and $v_{{\rm imp}} \approx 15 - 20\,{\rm cm/s}$. A more detailed parameter survey (Fig.\,\ref{b-a-of-largest-remnant-q=1.0-phid=40}b) shows that impacts with $v_{{\rm imp}}\sin \theta_{{\rm imp}} \approx 5.1\,{\rm cm/s}$ form EERs. Here, all largest remnants with $b/a < 0.3$ have the mass of $\approx M_{{\rm t}}+M_{{\rm i}}$, which means almost complete merging.

\cite{Sugiura-et-al2018} showed that the formation of elongated bodies needs large shear velocity $v_{{\rm imp}}\sin \theta_{{\rm imp}}$, which induces elongation of planetesimals. However, impacts with shear velocity that is too large result in splitting of colliding bodies, which do not lead to the formation of EERs. A moderate shear velocity $v_{{\rm imp}}\sin \theta_{{\rm imp}} \approx 5.1\,{\rm cm/s}$ results in sufficient elongation and merging to form EERs. Even if $v_{{\rm imp}}\sin \theta_{{\rm imp}} \approx 5.1\,{\rm cm/s}$, impacts with $v_{{\rm imp}} \geq 25\,{\rm cm/s}$ mainly result in catastrophic destruction, and those with $\theta_{{\rm imp}} \geq 20^{\circ}$ mainly result in hit-and-run collisions; neither of them are likely to form elongated objects. Therefore, we find that impacts with $\theta_{{\rm imp}} \leq 20^{\circ}$, $v_{{\rm imp}} \leq 25\,{\rm m/s}$ and $v_{{\rm imp}}\sin \theta_{{\rm imp}} \approx 5.1\,{\rm cm/s}$ form EERs for $q=1$ and $\phi_{{\rm d}}=40^{\circ}$.

\begin{figure}[!htb]
  \begin{center}
    \includegraphics[bb=0 0 960 498, width=1.0\linewidth,clip]{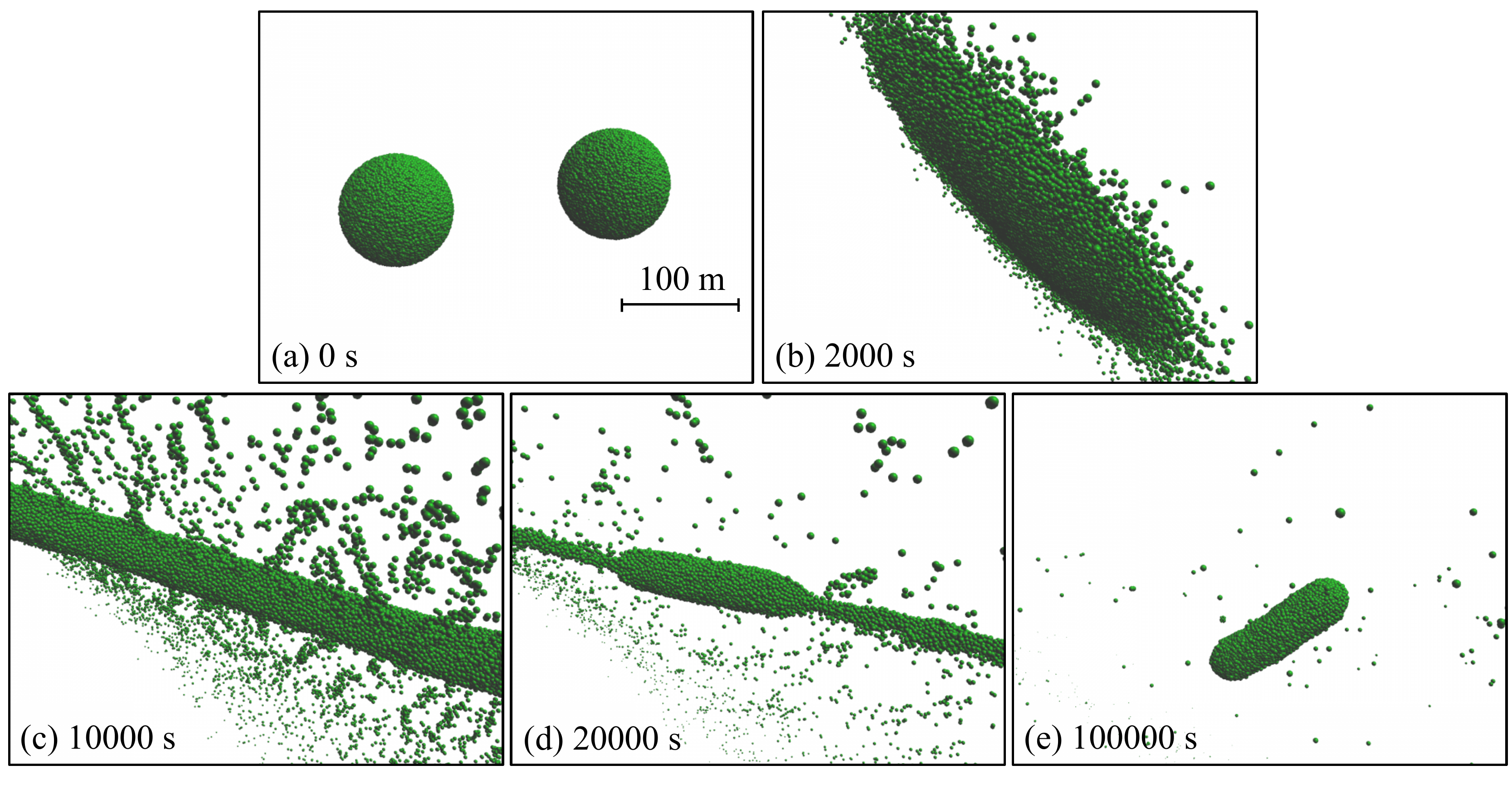}
    \caption{Snapshots of the third largest remnant formed through the impact with $q=1$, $\theta_{{\rm imp}}=15^{\circ}$, $v_{{\rm imp}}=24\,{\rm cm/s}$, and $\phi_{{\rm d}}=40^{\circ}$.}
    \label{subsequent-pictures-50m-1.0-15-24cms-phid=40}
  \end{center}
\end{figure}

We find one example of EER formation in smaller remnants. We obtain an EER that is formed from the third largest remnant for $\theta_{{\rm imp}}=15^{\circ}$ and $v_{{\rm imp}}=24\,{\rm cm/s}$ as shown in Fig\,\ref{subsequent-pictures-50m-1.0-15-24cms-phid=40}. This impact has a larger impact velocity than that shown in Fig.\,\ref{subsequent-pictures-50m-1.0-15-20cms-phid=40} and produces a large ejecta curtain (Fig.\,\ref{subsequent-pictures-50m-1.0-15-24cms-phid=40}b) which develops into a filamentary structure through reaccumulation of fragments in the direction perpendicular to the impact velocity vector (Fig.\,\ref{subsequent-pictures-50m-1.0-15-24cms-phid=40}c). Gravitational fragmentation of the filament leads to the formation of an EER (Fig.\,\ref{subsequent-pictures-50m-1.0-15-24cms-phid=40}d,e), which has the mass of $0.24M_{{\rm t}}$ and $b/a$ of 0.27. However, among the impact simulations with $\phi_{{\rm d}}=40^{\circ}$, this is the only example producing an EER that is not the largest remnant.

\subsection{Dependence on the friction angle}

\begin{figure}[!htb]
  \begin{center}
    \includegraphics[bb=0 0 859 540, width=0.8\linewidth,clip]{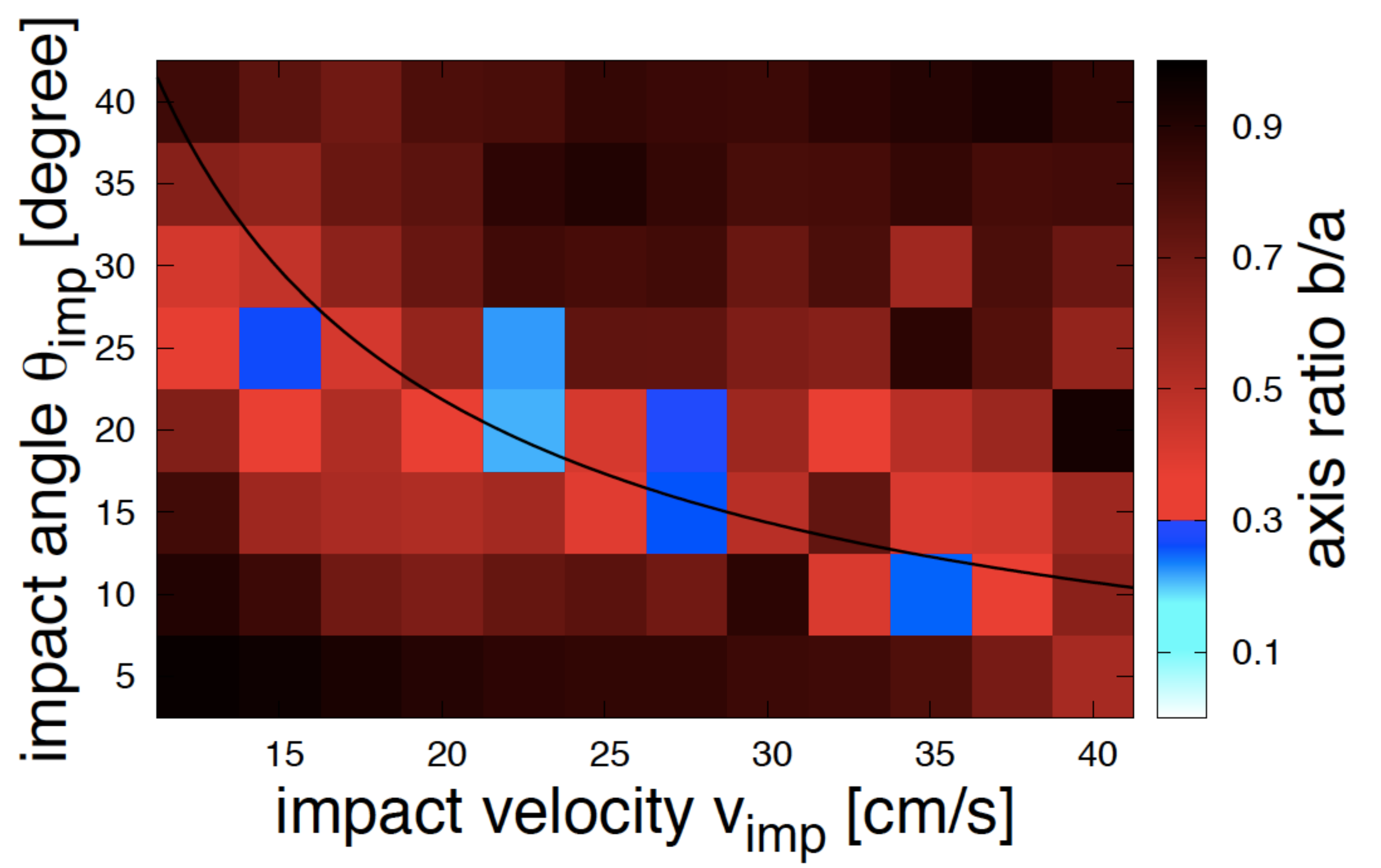}
    \caption{$b/a$ of the largest remnants resulting from various impact simulations with $q=1$ and $\phi_{{\rm d}}=50^{\circ}$. We vary parameters within $12.5\,{\rm cm/s} \leq v_{{\rm imp}} \leq 40\,{\rm cm/s}$ and $5^{\circ} \leq \theta_{{\rm imp}} \leq 40^{\circ}$ with increments $2.5\,{\rm cm/s}$ and $5^{\circ}$. Black curve shows $v_{{\rm imp}}\sin \theta_{{\rm imp}} = 7.4\,{\rm cm/s}$.}
    \label{b-a-of-largest-remnant-q=1.0-phid=50}
  \end{center}
\end{figure}

Figure \ref{b-a-of-largest-remnant-q=1.0-phid=50} represents $b/a$ of the largest remnants formed through various impacts with $q=1$ and $\phi_{{\rm d}}=50^{\circ}$. The obtained $b/a$ is not distributed smoothly in the $v_{{\rm imp}}$ and $\theta_{{\rm imp}}$ space. This non-smoothness might be due to the limited number of data points in this diagram compared to that in Fig.\,\ref{b-a-of-largest-remnant-q=1.0-phid=40}b. The formation of EERs occurs in a wider parameter space than that for $\phi_{{\rm d}}=40^{\circ}$. We find that parameters producing EERs concentrate around $v_{{\rm imp}} \sin \theta_{{\rm imp}} \approx 7.4\,{\rm cm/s}$ with $\theta_{{\rm imp}} \leq 30^{\circ}$ and $v_{{\rm imp}} \leq 35\,{\rm m/s}$. Note that for $\phi_{{\rm d}}=50^{\circ}$ EERs are also formed from smaller remnants in four parameters with $\theta_{{\rm imp}}\approx 25^{\circ}$ and $v_{{\rm imp}}\approx 20 - 30\,{\rm cm/s}$.

We also conduct impact simulations with $q=1$, $\phi_{{\rm d}}=30^{\circ}$, and various parameters within $12.5\,{\rm cm/s} \leq v_{{\rm imp}} \leq 35\,{\rm cm/s}$ and $5^{\circ} \leq \theta_{{\rm imp}} \leq 20^{\circ}$. Although the ranges of impact velocities and angles are almost the same as those in Fig.\,\ref{b-a-of-largest-remnant-q=1.0-phid=40} or Fig.\,\ref{b-a-of-largest-remnant-q=1.0-phid=50}, no EERs are formed through impacts with $\phi_{{\rm d}}=30^{\circ}$. Thus we find that the formation of EERs needs the friction angle $\phi_{{\rm d}} \geq 40^{\circ}$.

EERs are formed through collisional elongation of planetesimals and reaccumulation of fragments along the long axis (see Fig.\,\ref{subsequent-pictures-50m-1.0-15-20cms-phid=40} or Fig.\,\ref{subsequent-pictures-50m-1.0-15-24cms-phid=40}). In the reaccumulation phase, the frictional energy dissipation is given by (\citealt{Sugiura-et-al2018})

\begin{align}
  &E_{{\rm dis}} = \pi R_{{\rm r}}^{2} L \mu_{{\rm d}} p_{{\rm c}}, \nonumber \\
  &p_{{\rm c}}=\frac{2}{3}\pi G \rho_{0}^{2}R_{{\rm r}}^{2}, \label{Edis-at-reaccumulation}
\end{align}

\noindent where $G$ is the gravitational constant, $L$ is the displacement of fragments on the surface of an EER, and $R_{{\rm r}}$ and $p_{{\rm c}}$ are the mean radius and the central pressure of this EER, respectively. On the other hand, total kinetic energy of the fragments accumulating for the formation of the EER is estimated as

\begin{equation}
  E_{{\rm kin}}=\frac{1}{2}M_{{\rm r}} v^{2}_{{\rm esc, r}},
  \label{Ekin-for-EER-at-reaccumulation}
\end{equation}

\noindent where $M_{{\rm r}}$ is the mass of this EER and $v_{{\rm esc,r}}$ is the surface escape velocity from the EER. Equating $E_{{\rm kin}}$ and $E_{{\rm dis}}$, we have

\begin{equation}
\frac{L}{R_{{\rm r}}} \approx \frac{1}{\mu_{{\rm d}}}.
\end{equation}

\noindent If the displacement is much greater than the remnant size, i.e, $L/R_{{\rm r}} \gg 1$, EERs are not formed because of large deformation in the reaccumulation phase. Thus the formation of EERs requires $\mu_{{\rm d}} \gtrsim 1$, or $\phi_{{\rm d}} \gtrsim 45^{\circ}$. Interestingly, this condition solely depends on $\mu_{{\rm d}}$ so that it is applicable to impact events with various spatial scales.

\subsection{Dependence on the mass ratio}

\begin{figure}[!htb]
  \begin{center}
    \includegraphics[bb=0 0 960 498, width=1.0\linewidth,clip]{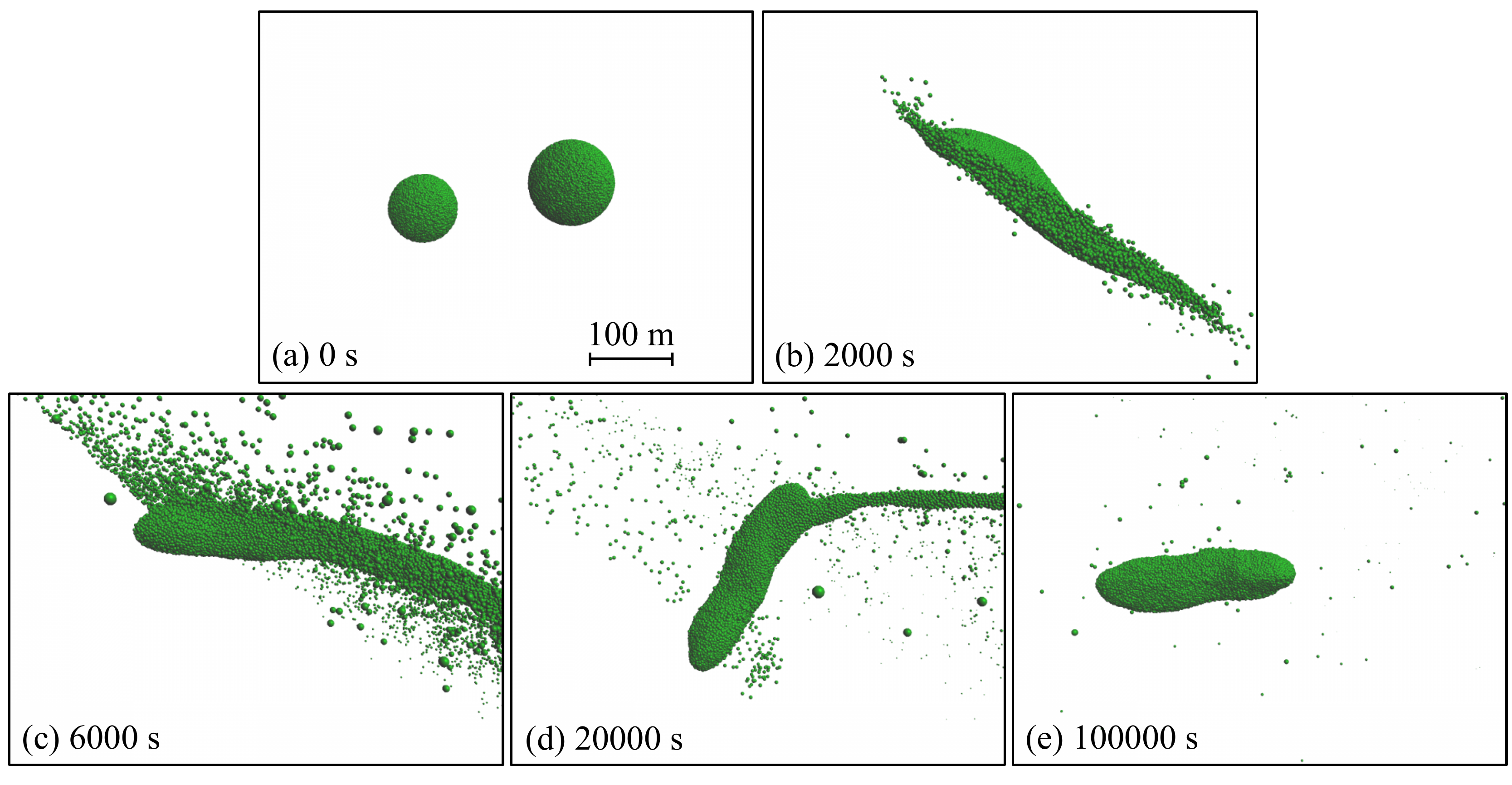}
    \caption{Snapshots of the largest remnant formed through the impact with $q=0.5$, $\theta_{{\rm imp}}=20^{\circ}$, $v_{{\rm imp}}=25\,{\rm cm/s}$, and $\phi_{{\rm d}}=50^{\circ}$.}
    \label{subsequent-pictures-50m-0.5-20-25cms-phid=50}
  \end{center}
\end{figure}

Figure \ref{subsequent-pictures-50m-0.5-20-25cms-phid=50} shows snapshots of the impact with $q=0.5$, $\phi_{{\rm d}}=50^{\circ}$, $\theta_{{\rm imp}}=20^{\circ}$, and $v_{{\rm imp}}=25\,{\rm cm/s}$. The asymmetric destruction of planetesimals occurs because of the unequal-mass impact (Fig.\,\ref{subsequent-pictures-50m-0.5-20-25cms-phid=50}b). The collision induces elongation (Fig.\,\ref{subsequent-pictures-50m-0.5-20-25cms-phid=50}c), which leads to the formation of an EER (Fig.\,\ref{subsequent-pictures-50m-0.5-20-25cms-phid=50}d,e). This remnant has $1.1M_{{\rm t}}$, $a=263\,{\rm m}$, and $b=72\,{\rm m}$, so that $b/a = 0.27$.

\begin{figure}[!htb]
  \begin{center}
    \includegraphics[bb=0 0 859 540, width=0.8\linewidth,clip]{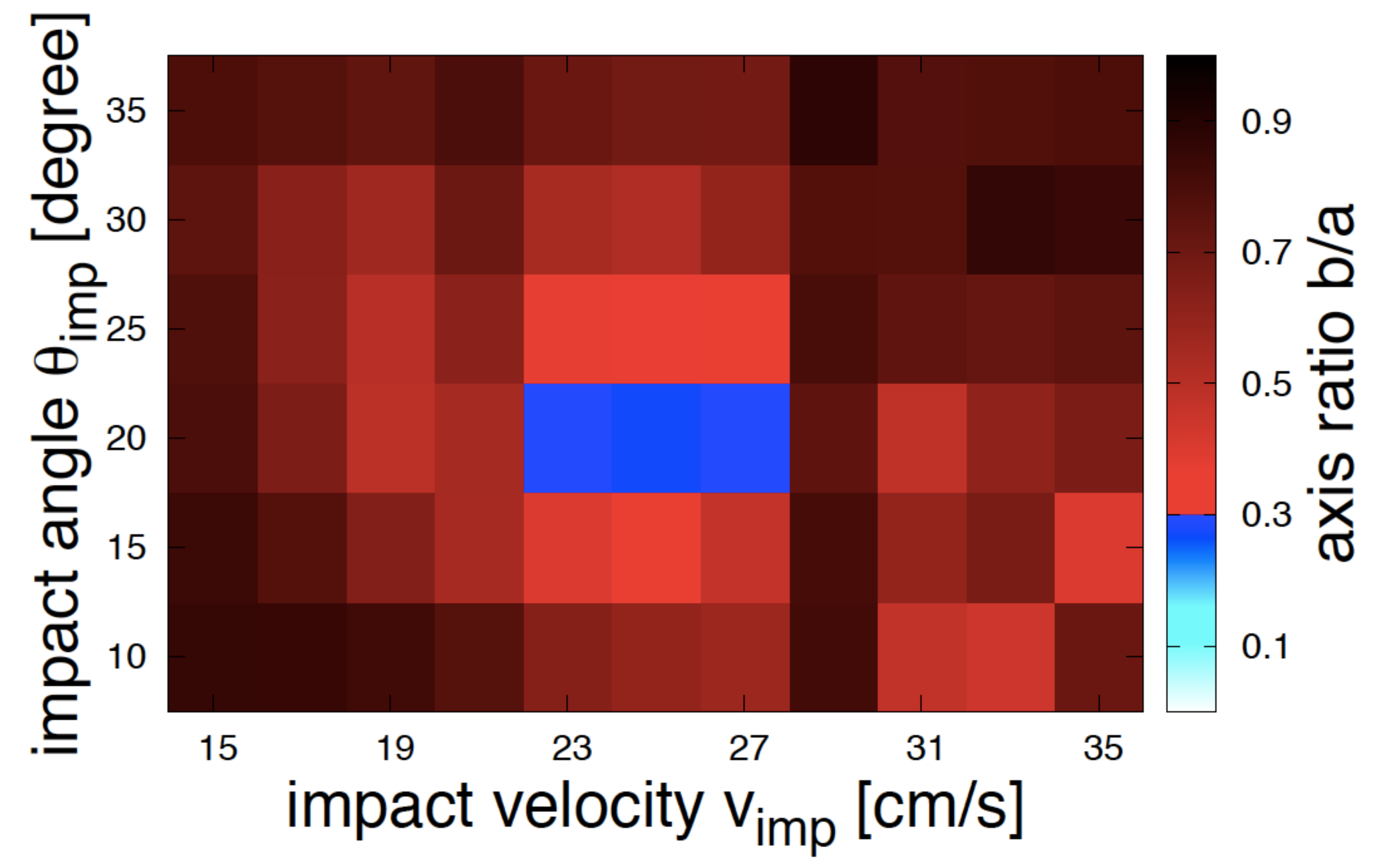}
    \caption{$b/a$ of the largest remnants resulting from various impacts with $q=0.5$ and $\phi_{{\rm d}}=50^{\circ}$. We vary parameters within $15\,{\rm cm/s} \leq v_{{\rm imp}} \leq 35\,{\rm cm/s}$ and $10^{\circ} \leq \theta_{{\rm imp}} \leq 35^{\circ}$ with increments $2\,{\rm cm/s}$ and $5^{\circ}$. }
    \label{b-a-of-largest-remnant-q=0.5-phid=50}
  \end{center}
\end{figure}

Figure \ref{b-a-of-largest-remnant-q=0.5-phid=50} represents $b/a$ of the largest remnants in various impact simulations with $q=0.5$ and $\phi_{{\rm d}}=50^{\circ}$. We find that three impacts with $\theta_{{\rm imp}}=20^{\circ}$ and $v_{{\rm imp}}\approx 25\,{\rm cm/s}$ form EERs.

We also conduct impact simulations with $q=0.25$, $\phi_{{\rm d}}=50^{\circ}$, and various parameters within $15\,{\rm cm/s} \leq v_{{\rm imp}} \leq 55\,{\rm cm/s}$ and $10^{\circ} \leq \theta_{{\rm imp}} \leq 35^{\circ}$. Although impacts in this parameter range include various types of collisions (merging, hit-and-run, and catastrophic destruction), no impacts with $q=0.25$ form EERs. Overall deformation of target planetesimal requires $q\sim 1$ because collisions with small impactors induce local deformation with the size comparable to that of impactors. Therefore the formation of EERs requires $q \geq 0.5$.

\section{Discussion}

In Figs\,\ref{b-a-of-largest-remnant-q=1.0-phid=40}, \ref{b-a-of-largest-remnant-q=1.0-phid=50}, and \ref{b-a-of-largest-remnant-q=0.5-phid=50}, the parameters forming EERs are limited and the collisional probability for EER formation is low. However, formation of extremely elongated objects through other processes is also difficult. For example, if four spherical planetesimals with equal size stick together in a completely straight line, an object with $b/a=0.25$ is formed. The collisional formation of the body requires three nearly head-on impacts with low velocities; such probability might be smaller than that of EER formation due to elongation through single collision. Thus, we focus on EER formation through collisional elongation, and estimate the dynamical environment for such collisions.

Impact velocities required to produce EERs ($15\,{\rm cm/s} \leq v_{{\rm imp}} \leq 40\,{\rm cm/s}$) are much smaller than typical orbital velocity around a star $\sim 10\,{\rm km/s}$. The velocity dispersion, which determines mean impact velocities, should be $10^{-5}$ times smaller than the orbital velocity. A dynamically cold protoplanetary disk is required in order to achieve such a low velocity dispersion. Here, we focus on turbulence and gravitational interaction with large bodies as dynamical excitation of planetesimals in protoplanetary disks, and give some constraints for environment such that $v_{{\rm imp}} \leq 40\,{\rm cm/s}$.

 It should be noted that the impact velocity to form extremely elongated objects may be increased if we additionally consider cohesion of granular material. The cohesion of lunar soil is estimated to be $1\,{\rm kPa}$ (\citealt{Heiken-et-al1991}). In contrast, central pressure of a planetesimal with the radius $50\,{\rm m}$ is about $10\,{\rm Pa}$, which gives the typical value of frictional stress. Thus impact velocity required for collisional elongation may become ten times larger if cohesion is included. Note that the fragments produced through high-velocity impacts tend to have high ejection velocities and easily escape from the system. On the other hand, cohesion is apt to prevent elongated shapes due to its large tensile strength if elongation occurs. The effect of cohesion will be investigated in our future papers. 

\subsection{Turbulence}
\begin{figure}[!htb]
  \begin{center}
    \includegraphics[bb=0 0 859 540, width=1.0\linewidth,clip]{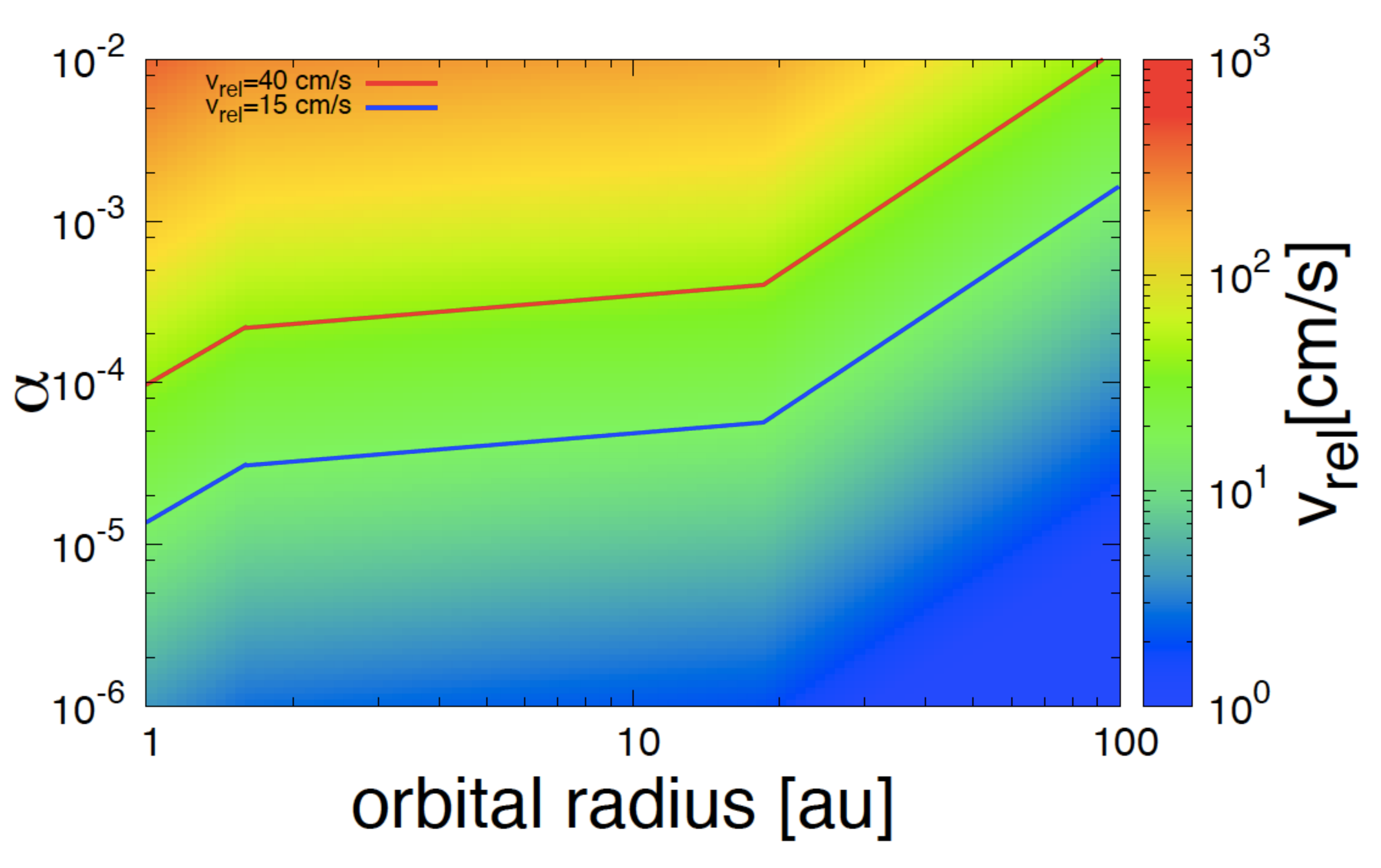}
    \caption{Relative velocity between two planetesimals with the radius $100\,{\rm m}$ induced by turbulence in a disk with the minimum mass solar nebula model. Horizontal axis shows the distance from the central star, and vertical axis shows the Shakura-Sunyaev $\alpha$ parameter. Red and blue solid curves show the contour of $v_{{\rm rel}}=40\,{\rm cm/s}$ and $v_{{\rm rel}}=15\,{\rm cm/s}$, respectively.}
    \label{vrel-turbulence-color-contour-for-MMSN-s=100m-v-lt-40cms}
  \end{center}
\end{figure}

 Turbulent stirring increases relative velocity $v_{{\rm rel}}$ between bodies in protoplanetary disks. For $100\,{\rm m}$ sized bodies, the stopping timescale $t_{{\rm s}}$ due to gas drag in the minimum mass solar nebula model with a central star with the solar mass and luminosity is given by (\citealt{Hayashi1981,Weidenschilling1977}) 

\begin{eqnarray}
  t_{{\rm s}} = \left\{ \begin{array}{lll}
    2.2\times 10^{1} \Bigl( \frac{r}{1\,{\rm AU}} \Bigr)^{3/2}\Omega_{{\rm K}}^{-1} & {\rm for} & r > 19\,{\rm AU}  \\
    3.6\times 10^{3}\Bigl( \frac{r}{1\,{\rm AU}} \Bigr)^{-1/4}\Omega_{{\rm K}}^{-1} & {\rm for} & 1.6\,{\rm AU} < r < 19\,{\rm AU} \\
    1.8\times 10^{3}\Bigl( \frac{r}{1\,{\rm AU}} \Bigr)^{5/4}\Omega_{{\rm K}}^{-1} & {\rm for} & r < 1.6\,{\rm AU} \\
  \end{array} \right. ,
  \label{stopping-time-vs-r}
\end{eqnarray}

\noindent where $r$ is the orbital radius and $\Omega_{{\rm K}}$ is the Keplerian angular frequency. The gas drag is determined by the Epstein law for $r>19\,{\rm AU}$, the Stokes law for $1.6\,{\rm AU}<r<19\,{\rm AU}$, and the Newton law (high Reynolds number) for $r<1.6\,{\rm AU}$. For $t_{{\rm s}}\Omega_{{\rm K}} \gg 1$, $v_{{\rm rel}}$ controlled by turbulent stirring is given by (\citealt{Cuzzi-et-al2001, Ormel-and-Cuzzi2007}) 

\begin{equation}
  v_{{\rm rel}}\approx C_{{\rm s}}\sqrt{3\alpha (\Omega_{{\rm K}}t_{{\rm s}})^{-1}},
  \label{vrel-turbulence}
\end{equation}

\noindent where $C_{{\rm s}}$ is the sound speed of gas and $\alpha$ is the strength of turbulence in Shakura-Sunyaev prescription. Figure \ref{vrel-turbulence-color-contour-for-MMSN-s=100m-v-lt-40cms} shows $v_{{\rm rel}}$ as a function of $r$ and $\alpha$, and we find that $\alpha < 5\times 10^{-4}$ for $1\,{\rm AU}<r<20\,{\rm AU}$ and $\alpha < 5\times 10^{-4}(r/20\,{\rm AU})^{2}$ for $r>20\,{\rm AU}$ are roughly required for $v_{{\rm imp}} \leq 40\,{\rm cm/s}$.

Magnetorotational instability (MRI) is a dominant source of turbulence. MRI is active in vicinities of a central star with high ionized fraction ($r<0.1\,{\rm AU}$) or sufficiently distant from a central star with low surface density of gas ($r>10\,{\rm AU}$), which results in strong turbulence with $\alpha \sim 10^{-2}$ (e.g.,\,\citealt{Flock-et-al2017}). MRI non-active regions (dead zones) are roughly considered to be the mid-plane with $0.1\,{\rm AU} < r < 10\,{\rm AU}$ (e.g.,\,\citealt{Kretke-et-al2009}). The strength of turbulence in dead zones is estimated to be $10^{-6} < \alpha < 10^{-3}$ (e.g.,\,\citealt{Flock-et-al2017, Ilgner-and-Nelson2008, Mori-et-al2017}). Thus impacts with $v_{{\rm imp}} \leq 40\,{\rm cm/s}$ between two $100\,{\rm m}$ sized planetesimals could occur in dead zones with $1\,{\rm AU} < r < 10\,{\rm AU}$.

On the other hand, observations of a protoplanetary disk suggest that $\alpha$ for the outer disk is smaller than values expected from MRI turbulence. \cite{Pinte-et-al2016} show that apparent gaps of HL Tau require $\alpha \sim 10^{-4}$. Note that such a small $\alpha$ is favorable for creating the observed multiple ring-like structure in HL Tau through secular gravitational instability (\citealt{Takahashi-and-Inutsuka2014, Takahashi-and-Inutsuka2016, Tominaga-et-al2018}). Impact velocities of $v_{{\rm imp}} \leq 40\,{\rm cm/s}$ are found in protoplanetary disks with $\alpha \sim 10^{-4}$ (see Fig.\,\ref{vrel-turbulence-color-contour-for-MMSN-s=100m-v-lt-40cms}), and thus 1I/`Oumuamua may have formed in a protoplanetary disk like HL Tau.

\subsection{Size of larger planetesimals}

Gravitational stirring by larger planetesimals increases the relative velocity between surrounding smaller planetesimals. We consider two groups of planetesimals: One of them is small planetesimals with the radius $100\,{\rm m}$, and the other one is large planetesimals with the radius $R$. Through viscous stirring by large planetesimals, the relative velocity of small planetesimals $u$ increases at a rate of (\citealt{Ida-and-Makino1993})

\begin{equation}
  \frac{1}{u}\left.\frac{du}{dt}\right|_{{\rm vs}} \sim \Omega_{{\rm K}}\frac{\Sigma}{\rho_{{\rm s}}R}\Bigl( \frac{v_{{\rm esc}}}{u} \Bigr)^{4},
  \label{viscous-stirring}
\end{equation}

\noindent where $\rho_{{\rm s}}$ is the density of bodies, $\Sigma$ is the surface density of large planetesimals, and $v_{{\rm esc}}$ is the escape velocity from a large planetesimal. The relative velocity is decreased through gas drag at a rate of (\citealt{Adachi-et-al1976, Kobayashi2015})

\begin{equation}
  \frac{1}{u}\left.\frac{du}{dt}\right|_{{\rm gd}} \sim -\frac{1}{t_{{\rm s}}},
  \label{gas-damping}
\end{equation}

\noindent which is inversely proportional to the stopping time. The balance between the increasing rate Eq.\,(\ref{viscous-stirring}) and the decreasing rate Eq.\,(\ref{gas-damping}) gives the relative velocity at a steady state

\begin{align}
  u &= \Bigl( \frac{t_{{\rm s}}\Omega_{{\rm K}}\Sigma}{\rho_{{\rm s}}R} \Bigr)^{1/4}v_{{\rm esc}}. \label{random-velocity-by-big-body}
\end{align}

\begin{figure}[!htb]
  \begin{center}
    \includegraphics[bb=0 0 859 540, width=1.0\linewidth,clip]{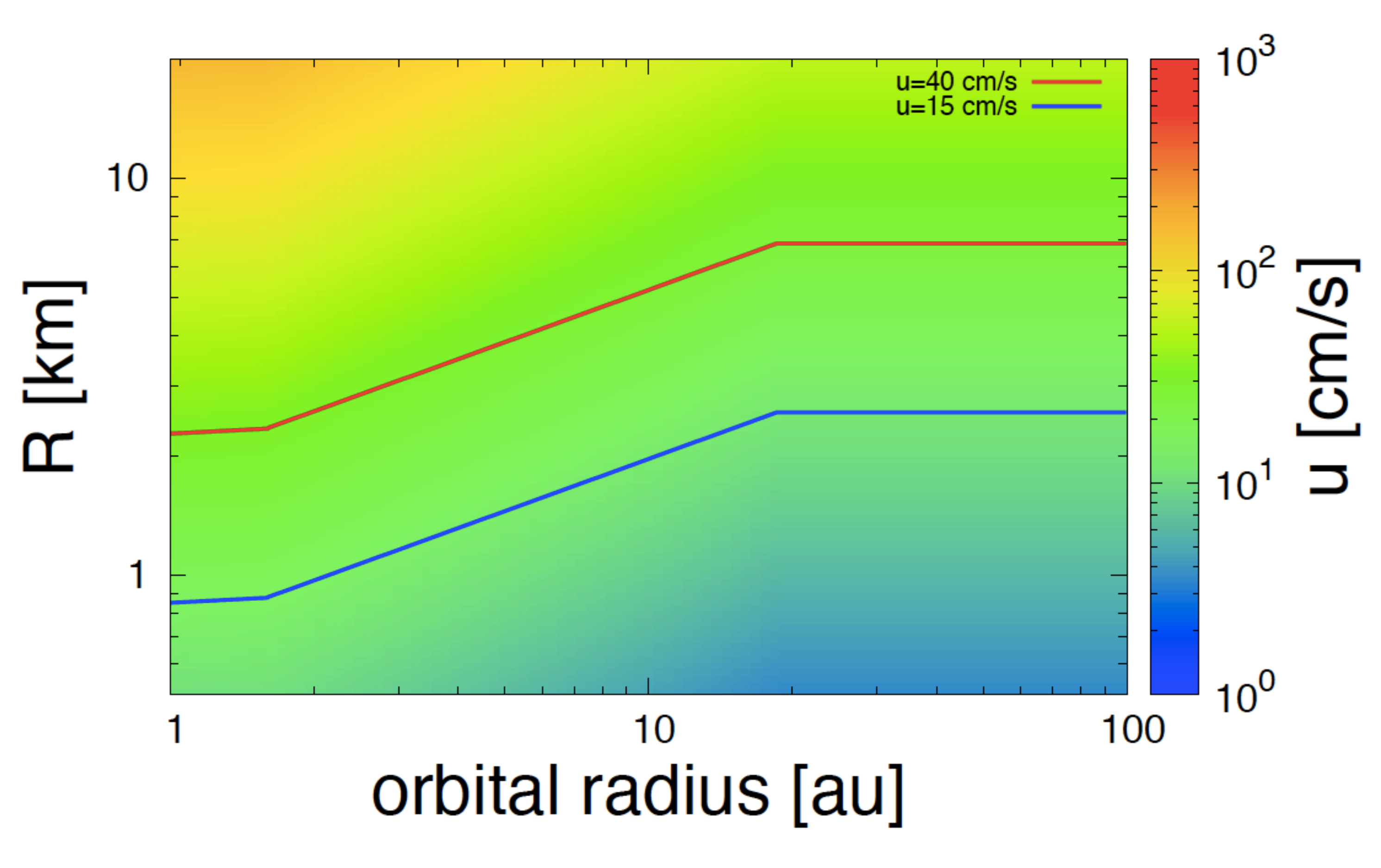}
    \caption{Relative velocity between planetesimals with the radius $100\,{\rm m}$ induced by large planetesimals with the radius $R$ in a disk with the minimum mass solar nebula model. Horizontal axis shows the distance from the central star, and vertical axis shows the radius of large planetesimals $R$. Red and blue solid curves show the contour of $u=40\,{\rm cm/s}$ and $u=15\,{\rm cm/s}$, respectively.}
    \label{vrel-large-planetesimals-color-contour-for-MMSN-s=100m-v-lt-40cms}
  \end{center}
\end{figure}

\noindent Figure \ref{vrel-large-planetesimals-color-contour-for-MMSN-s=100m-v-lt-40cms} shows the relative velocity in the minimum mass solar nebula. The surface density of large planetesimals, $\Sigma$, is set at 0.01 of the total solid surface density given by \cite{Hayashi1981}. From Fig.\,\ref{vrel-large-planetesimals-color-contour-for-MMSN-s=100m-v-lt-40cms}, we find that $R<2\,{\rm km}$ for $r \approx 1\,{\rm AU}$ and $R<7\,{\rm km}$ for $r>20\,{\rm AU}$ are required to achieve $u<40\,{\rm cm/s}$.

1I/`Oumuamua is considered to have been ejected from another planetary system. Ejection from a planetary system requires strong gravitational scattering by massive objects such as gas giants, passing stars, and central binary stars. However, such gravitational perturbation significantly increases the relative velocity between planetesimals. Thus the shape of 1I/`Oumuamua is probably formed in a place without large bodies, which may be in an outer protoplanetary disk, and then 1I/`Oumuamua is drifted inward to the vicinity of gas giants or central binary stars where it could be ejected (e.g.,\,\citealt{Raymond-et-al2017, Jackson-et-al2017}). Ejection by central binary stars may be favorable because it could explain lack of cometary activity from 1I/`Oumuamua. A close encounter with binary stars could evaporate the volatiles from 1I/`Oumuamua (\citealt{Jackson-et-al2017}). In contrast, an early stellar encounter ejects planetesimals directly from an outer disk (e.g.,\,\citealt{Kobayashi-et-al2005}).

Gravitational scattering of planetesimals due to gas giants is supposed to produce the Oort cloud (e.g.,\,\citealt{Bandermann-and-Wolstencroft1971, Fernandez1978, Higuchi-et-al2006}). We do not discard the possibility that 1I/`Oumuamua may come from the Oort cloud, since the extremely elongated shape of 1I/`Oumuamua could be formed in the primordial solar system through collisional elongation. Thus, 1I/`Oumuamua might be formed in the primordial solar system, scattered by gas giants into the Oort cloud, and then perturbed again by possibly a stellar encounter achieving the eccentricity larger than unity.

\subsection{Survivability of extremely elongated shapes through ejection processes}
As we discussed in the previous subsection, 1I/`Oumuamua is considered to have been ejected from another planetary system. However, the ejection requires a close encounter with a gas giant or a star, and the extremely elongated shape of 1I/`Oumuamua may possibly be destroyed due to tidal force if the encounter distance is too small. Here, we discuss the survivability of the shape of 1I/`Oumuamua through ejection processes.

We consider a prolate object with major axis length $a$, intermediate (minor) axis length $b$, and density $\rho_{{\rm s}}$ orbiting around a central body with density $\rho_{{\rm s}}$, radius $R_{{\rm c}}$, and mass $M_{{\rm c}}=(4/3)\pi R_{{\rm c}}^{3}\rho_{{\rm s}}$. The prolate body is efficiently deformed when the major axis points toward the central body, and the specific tidal force acting on the surface of the prolate object is

\begin{equation}
  F_{{\rm t}}=\frac{GM_{{\rm c}}a}{d^{3}},
  \label{tidal-force-on-prolate-body}
\end{equation}

\noindent where $d$ is the distance between the prolate object and the central body. On the other hand, the specific self-gravitational force of the prolate object on the surface on its major axis is given by (\citealt{Chandrasekhar1969})

\begin{equation}
  F_{{\rm s}}=\pi G \rho_{{\rm s}}A_{1}a,
  \label{self-gravtational-force-on-plorate-body}
\end{equation}

\noindent where

\begin{align}
  & A_{1}=\frac{1-e^{2}}{e^{3}} \ln \Bigl( \frac{1+e}{1-e} \Bigr) -2 \frac{1-e^{2}}{e^{2}}, \nonumber \\
  & e=\sqrt{1-\frac{b^{2}}{a^{2}}}. \label{eq-for-self-gravity-on-prolate-body}
\end{align}

\noindent The tidal deformation or disruption occurs when $F_{{\rm t}} > F_{{\rm s}}$. The condition is written as

\begin{equation}
  d < d_{{\rm R}} \equiv R_{{\rm c}}\sqrt[3]{\frac{4}{3}A_{1}^{-1}}.
  \label{Roche-radius-for-prolate-object}
\end{equation}

\noindent For an extreme shape with $b/a=0.1$, $d_{{\rm R}}=3.2R_{{\rm c}}$, which is comparable to the radius of the central body.

Planetesimals are ejected due to interactions with binary stars once their orbital semi-major axes are smaller than $a_{{\rm c}}$, which is estimated to be $a_{{\rm c}}\approx 2.4 a_{{\rm b}}$ for an equal-mass circular-orbit binary with orbital separation $a_{{\rm b}}$ (\citealt{Holman-and-Wiegert1999}). The binary separation $a_{{\rm b}}$ is much larger than $d_{{\rm R}}$; planetesimals are ejected before they experience tidal disruption.

Planetesimals are ejected from the parent disk due to a stellar encounter with the encounter distance of $\sim 100\,{\rm AU}$ (\citealt{Kobayashi-and-Ida2001}). Therefore, the encounter distance is much larger than $d_{{\rm R}}$ so that tidal deformation is negligible.

A giant planet perturbs a circular-orbit planetesimal, resulting in an increase in orbital eccentricity of the planetesimal (\citealt{Ohtsuki-et-al2002}). Its eccentricity becomes larger than $\sim 0.3$, and the planetesimal may then escape from the planetary system by the scattering due to the gas giant (\citealt{Higuchi-et-al2006}). The encounter distance required for the eccentricity increase and the escape is estimated to be on the order of the Hill radius (\citealt{Ida1990, Fernandez1978}), which is 740 times larger than the radius of the planet for Jupiter. Therefore, the extremely elongated shape of 1I/`Oumuamua can be kept through ejection processes.

\section{Conclusion}

1I/`Oumuamua is the first interstellar object observed. This object may have an extremely elongated shape with the ratio between intermediate axis length $b$ and major axis length $a$ less than 0.3. Clarifying the mechanism or necessary conditions to form extremely elongated objects may reveal the environment of the planetary system that 1I/`Oumuamua originated from.

We conduct numerical simulations of planetesimal collisions using SPH for elastic dynamics with the friction model and the self-gravity. We assume that planetesimals do not have tensile strength but have shear strength and shear strength is determined by the friction of granular material. Collisions for $50\,{\rm m}$ sized planetesimals are simulated with various ratios of impactor mass to target mass $q$, friction angles of granular material $\phi_{{\rm d}}$, impact velocities $v_{{\rm imp}}$, and impact angles $\theta_{{\rm imp}}$. As a result, we confirm that extremely elongated shapes of impact outcomes with $b/a<0.3$ are formed through some impacts with the following conditions:

\begin{itemize}
\item $v_{{\rm imp}}\sin \theta_{{\rm imp}} \approx 5.1\,{\rm cm/s}$, $\theta_{{\rm imp}} \leq 20^{\circ}$, and $v_{{\rm imp}} \leq 24\,{\rm cm/s}$ for $q=1$ and $\phi_{{\rm d}}=40^{\circ}$ (Fig.\,\ref{b-a-of-largest-remnant-q=1.0-phid=40}).
\item $v_{{\rm imp}}\sin \theta_{{\rm imp}} \approx 7.4\,{\rm cm/s}$, $\theta_{{\rm imp}} \leq 30^{\circ}$, and $v_{{\rm imp}} \leq 35\,{\rm cm/s}$ for $q=1$ and $\phi_{{\rm d}}=50^{\circ}$ (Fig.\,\ref{b-a-of-largest-remnant-q=1.0-phid=50}).
\item $\theta_{{\rm imp}} = 20^{\circ}$ and $23\,{\rm cm/s} \leq v_{{\rm imp}} \leq 27\,{\rm cm/s}$ for $q=0.5$ and $\phi_{{\rm d}}=50^{\circ}$ (Fig.\,\ref{b-a-of-largest-remnant-q=0.5-phid=50}).
\end{itemize}

\noindent Thus the formation of extremely elongated objects through collisions roughly requires $q \geq 0.5$, $\phi_{{\rm d}} \geq 40^{\circ}$, $v_{{\rm imp}} \leq 40\,{\rm cm/s}$, and $\theta_{{\rm imp}} \leq 30^{\circ}$.

We estimate the conditions to produce 1I/`Oumuamua, i.e., the conditions for impacts with $v_{{\rm imp}}<40\,{\rm cm/s}$ between $100\,{\rm m}$ sized bodies, in a protoplanetary disk given by the minimum mass solar nebula model. Turbulent stirring increases collisional velocities, and impacts with $v_{{\rm imp}}<40\,{\rm cm/s}$ require $\alpha < 5\times 10^{-4}$ for $1\,{\rm AU}<r<20\,{\rm AU}$ and $\alpha < 5\times 10^{-4}(r/20\,{\rm AU})^{2}$ for $r>20\,{\rm AU}$, where $\alpha$ is the turbulent strength and $r$ is the orbital radius of colliding bodies. These values of $\alpha$ are found in MRI inactive disks such as HL Tau. Large planetesimals also increase relative velocities between small planetesimals because of gravitational stirring. We estimate that planetesimals with radius $R > 7\,{\rm km}$ induce $v_{{\rm imp}} > 40\,{\rm cm/s}$. 

Some studies suggest other mechanisms to form extremely elongated objects such as catastrophic disruption, tidal disruption, particle abrasion, and deformation due to spin-up. To constrain the mechanism to form extremely elongated objects, detailed observation of asteroids with such shapes are useful. The Near Earth Asteroid (1865) Cerberus also shows the large luminosity variation with $\approx 2.3\,{\rm mag}$ and may have a shape with $b/a \approx 0.33$ (\citealt{Durech-et-al2012}). Information such as detailed shapes or density of extremely elongated asteroids will give us clues to reveal the formation mechanism.

In this paper, we provide quantitative conditions to form extremely elongated objects through collisions between granular planetesimals. Other mechanisms such as tidal elongation and/or catastrophic disruption and gravitational reaccumulation may form extremely elongated objects, which can also be investigated with our numerical code. Therefore, we will further explore the detailed conditions to form extremely elongated objects through other mechanisms in future work, which will constrain the origin of such shapes in more detail.

\section*{Acknowledgement}
KS is supported by JSPS KAKENHI Grant Number JP 17J01703. HK and SI are supported by Grant-in-Aid for Scientific Research (18H05436, 18H05437, 18H05438, 17K05632, 17H01105, 17H01103, 23244027, 16H02160) and by JSPS Core-to-Core Program ``International Network of Planetary Science''. Numerical simulations in this work were carried out on Cray XC30 at Center for Computational Astrophysics, National Astronomical Observatory of Japan. We thank anonymous reviewers for valuable and detailed comments on our manuscript.

\section*{References}

\bibliography{mybibfile}

\begin{thebibliography}{}

\bibitem[\protect\citename{{Adachi} {\em et~al.\ }\relax,
  }1976]{Adachi-et-al1976}
{Adachi}, I., {Hayashi}, C., \& {Nakazawa}, K. 1976.
\newblock {The gas drag effect on the elliptical motion of a solid body in the
  primordial solar nebula.}
\newblock {\em Prog. Theor. Phys.}, {\bf 56}(Dec.), 1756--1771.

\bibitem[\protect\citename{{Bandermann} \& {Wolstencroft},
  }1971]{Bandermann-and-Wolstencroft1971}
{Bandermann}, L.~W., \& {Wolstencroft}, R.~D. 1971.
\newblock {Energy changes in close planetary encounters}.
\newblock {\em Mon. Not. R. Astron. Soc.}, {\bf 152}, 377.

\bibitem[\protect\citename{{Bannister} {\em et~al.\ }\relax,
  }2017]{Bannister-et-al2017}
{Bannister}, M.~T., {Schwamb}, M.~E., {Fraser}, W.~C., {Marsset}, M.,
  {Fitzsimmons}, A., {Benecchi}, S.~D., {Lacerda}, P., {Pike}, R.~E.,
  {Kavelaars}, J.~J., {Smith}, A.~B., {Stewart}, S.~O., {Wang}, S.-Y., \&
  {Lehner}, M.~J. 2017.
\newblock {{Col-OSSOS}: {C}olors of the Interstellar Planetesimal
  {1I}/{'O}umuamua}.
\newblock {\em Astrophys. J.}, {\bf 851}(Dec.), L38.

\bibitem[\protect\citename{Benz \& Asphaug, }1995]{Benz-and-Asphaug1995}
Benz, W., \& Asphaug, E. 1995.
\newblock Simulations of brittle solids using smooth particle hydrodynamics.
\newblock {\em Comput. Phys. Commun.}, {\bf 87}, 253--265.

\bibitem[\protect\citename{Benz \& Asphaug, }1999]{Benz-and-Asphaug1999}
Benz, W., \& Asphaug, E. 1999.
\newblock Catastrophic disruption revisited.
\newblock {\em Icarus}, {\bf 142}, 5--20.

\bibitem[\protect\citename{{Bolin} {\em et~al.\ }\relax,
  }2018]{Bolin-et-al2017}
{Bolin}, B.~T., {Weaver}, H.~A., {Fernandez}, Y.~R., {Lisse}, C.~M.,
  {Huppenkothen}, D., {Jones}, R.~L., {Juri{\'c}}, M., {Moeyens}, J.,
  {Schambeau}, C.~A., {Slater}, C.~T., {Ivezi{\'c}}, {\v Z}., \& {Connolly},
  A.~J. 2018.
\newblock {{APO} Time-resolved Color Photometry of Highly Elongated
  Interstellar Object {1I}/{'O}umuamua}.
\newblock {\em Astrophys. J.}, {\bf 852}(Jan.), L2.

\bibitem[\protect\citename{{Bottke} {\em et~al.\ }\relax,
  }2005]{Bottke-et-al2005}
{Bottke}, W.~F., {Durda}, D.~D., {Nesvorn{\'y}}, D., {Jedicke}, R.,
  {Morbidelli}, A., {Vokrouhlick{\'y}}, D., \& {Levison}, H. 2005.
\newblock {The fossilized size distribution of the main asteroid belt}.
\newblock {\em Icarus}, {\bf 175}(May), 111--140.

\bibitem[\protect\citename{{Chandrasekhar}, }1969]{Chandrasekhar1969}
{Chandrasekhar}, S. 1969.
\newblock {\em {Ellipsoidal figures of equilibrium}}.

\bibitem[\protect\citename{{{\'C}uk}, }2018]{Cuk2017}
{{\'C}uk}, M. 2018.
\newblock {{1I}/{'Oumuamua} as a Tidal Disruption Fragment from a Binary Star
  System}.
\newblock {\em Astrophys. J.}, {\bf 852}(Jan.), L15.

\bibitem[\protect\citename{{Cuzzi} {\em et~al.\ }\relax,
  }2001]{Cuzzi-et-al2001}
{Cuzzi}, J.~N., {Hogan}, R.~C., {Paque}, J.~M., \& {Dobrovolskis}, A.~R. 2001.
\newblock {Size-selective Concentration of Chondrules and Other Small Particles
  in Protoplanetary Nebula Turbulence}.
\newblock {\em Astrophys. J.}, {\bf 546}(Jan.), 496--508.

\bibitem[\protect\citename{{de la Fuente Marcos} \& {de la Fuente Marcos},
  }2017]{de-la-Fuente-Marcos-double2017}
{de la Fuente Marcos}, C., \& {de la Fuente Marcos}, R. 2017.
\newblock {Pole, Pericenter, and Nodes of the Interstellar Minor Body {A/2017
  U1}}.
\newblock {\em Res. Notes AAS}, {\bf 1}(Dec.), 5.

\bibitem[\protect\citename{Domokos {\em et~al.\ }\relax,
  }2009]{Domokos-et-al2009}
Domokos, G., Sipos, A.~{\'A}., Szab{\'o}, Gy.~M., \& V{\'a}rkonyi, P.~L. 2009.
\newblock Formation of Sharp Edges and Planar Areas of Asteroids by Polyhedral
  Abrasion.
\newblock {\em Astrophys. J.}, {\bf 699}(1), L13.

\bibitem[\protect\citename{Domokos {\em et~al.\ }\relax,
  }2017]{Domokos-et-al2017}
Domokos, G., Sipos, A.~{\'A}., Szab{\'o}, G.~M., \& V{\'a}rkonyi, P.~L. 2017.
\newblock Explaining the Elongated Shape of {'Oumuamua} by the Eikonal Abrasion
  Model.
\newblock {\em Res. Notes AAS}, {\bf 1}(1), 50.

\bibitem[\protect\citename{{Drahus} {\em et~al.\ }\relax,
  }2018]{Drahus-et-al2018}
{Drahus}, M., {Guzik}, P., {Waniak}, W., {Handzlik}, B., {Kurowski}, S., \&
  {Xu}, S. 2018.
\newblock {Tumbling motion of 1I/`Oumuamua and its implications for the body's
  distant past}.
\newblock {\em Nature Astron.}, {\bf 2}(May), 407--412.

\bibitem[\protect\citename{{Fernandez}, }1978]{Fernandez1978}
{Fernandez}, J.~A. 1978.
\newblock {Mass removed by the outer planets in the early solar system}.
\newblock {\em Icarus}, {\bf 34}(Apr.), 173--181.

\bibitem[\protect\citename{{Flock} {\em et~al.\ }\relax,
  }2017]{Flock-et-al2017}
{Flock}, M., {Fromang}, S., {Turner}, N.~J., \& {Benisty}, M. 2017.
\newblock {{3D} Radiation Nonideal Magnetohydrodynamical Simulations of the
  Inner Rim in Protoplanetary Disks}.
\newblock {\em Astrophys. J.}, {\bf 835}(Feb.), 230.

\bibitem[\protect\citename{{Fraser} {\em et~al.\ }\relax,
  }2018]{Fraser-et-al2017}
{Fraser}, W.~C., {Pravec}, P., {Fitzsimmons}, A., {Lacerda}, P., {Bannister},
  M.~T., {Snodgrass}, C., \& {Smoli'c}, I. 2018.
\newblock {The tumbling rotational state of {1I/`Oumuamua}}.
\newblock {\em Nature Astron.}, 2.

\bibitem[\protect\citename{{Hayashi}, }1981]{Hayashi1981}
{Hayashi}, C. 1981.
\newblock {Structure of the Solar Nebula, Growth and Decay of Magnetic Fields
  and Effects of Magnetic and Turbulent Viscosities on the Nebula}.
\newblock {\em Prog. Theor. Phys. Suppl.}, {\bf 70}, 35--53.

\bibitem[\protect\citename{{Heiken} {\em et~al.\ }\relax,
  }1991]{Heiken-et-al1991}
{Heiken}, G.~H., {Vaniman}, D.~T., \& {French}, B.~M. 1991.
\newblock {\em {Lunar sourcebook - {A} user's guide to the moon}}.

\bibitem[\protect\citename{{Higuchi} {\em et~al.\ }\relax,
  }2006]{Higuchi-et-al2006}
{Higuchi}, A., {Kokubo}, E., \& {Mukai}, T. 2006.
\newblock {Scattering of Planetesimals by a Planet: Formation of Comet Cloud
  Candidates}.
\newblock {\em Astron. J.}, {\bf 131}(Feb.), 1119--1129.

\bibitem[\protect\citename{{Holman} \& {Wiegert},
  }1999]{Holman-and-Wiegert1999}
{Holman}, M.~J., \& {Wiegert}, P.~A. 1999.
\newblock {Long-Term Stability of Planets in Binary Systems}.
\newblock {\em Astron. J.}, {\bf 117}(Jan.), 621--628.

\bibitem[\protect\citename{{Ida}, }1990]{Ida1990}
{Ida}, S. 1990.
\newblock {Stirring and dynamical friction rates of planetesimals in the solar
  gravitational field}.
\newblock {\em Icarus}, {\bf 88}(Nov.), 129--145.

\bibitem[\protect\citename{{Ida} \& {Makino}, }1993]{Ida-and-Makino1993}
{Ida}, S., \& {Makino}, J. 1993.
\newblock {Scattering of planetesimals by a protoplanet - Slowing down of
  runaway growth}.
\newblock {\em Icarus}, {\bf 106}(Nov.), 210.

\bibitem[\protect\citename{{Ilgner} \& {Nelson}, }2008]{Ilgner-and-Nelson2008}
{Ilgner}, M., \& {Nelson}, R.~P. 2008.
\newblock {Turbulent transport and its effect on the dead zone in
  protoplanetary discs}.
\newblock {\em Astron. Astrophys.}, {\bf 483}(June), 815--830.

\bibitem[\protect\citename{Iwasawa {\em et~al.\ }\relax,
  }2015]{Iwasawa-et-al2015}
Iwasawa, M., Tanikawa, A., Hosono, N., Nitadori, K., Muranushi, T., \& Makino,
  J. 2015.
\newblock {FDPS}: {A} Novel Framework for Developing High-performance Particle
  Simulation Codes for Distributed-memory Systems.
\newblock {\em Pages  1:1--1:10 of:} {\em Proceedings of the 5th International
  Workshop on Domain-Specific Languages and High-Level Frameworks for High
  Performance Computing}.
\newblock WOLFHPC '15.
\newblock New York, NY, USA: ACM.

\bibitem[\protect\citename{Iwasawa {\em et~al.\ }\relax,
  }2016]{Iwasawa-et-al2016}
Iwasawa, M., Tanikawa, A., Hosono, N., Nitadori, K., Muranushi, T., \& Makino,
  J. 2016.
\newblock Implementation and performance of {FDPS}: {A} framework for
  developing parallel particle simulation codes.
\newblock {\em Publ. Astron. Soc. Jpn.}, {\bf 68 (4)}, 54 (1--22).

\bibitem[\protect\citename{{Jackson} {\em et~al.\ }\relax,
  }2018]{Jackson-et-al2017}
{Jackson}, A.~P., {Tamayo}, D., {Hammond}, N., {Ali-Dib}, M., \& {Rein}, H.
  2018.
\newblock {Ejection of rocky and icy material from binary star systems:
  {I}mplications for the origin and composition of {1I/`Oumuamua}}.
\newblock {\em Mon. Not. R. Astron. Soc.}, Mar.

\bibitem[\protect\citename{Jewitt {\em et~al.\ }\relax,
  }2017]{Jewitt-et-al2017}
Jewitt, D., Luu, J., Rajagopal, J., Kotulla, R., Ridgway, S., Liu, W., \&
  Augusteijn, T. 2017.
\newblock Interstellar Interloper {1I/2017 U1}: Observations from the {NOT} and
  {WIYN} Telescopes.
\newblock {\em Astrophys. J.}, {\bf 850}(2), L36.

\bibitem[\protect\citename{Jutzi, }2015]{Jutzi2015}
Jutzi, M. 2015.
\newblock SPH calculations of asteroid disruptions: {T}he role of pressure
  dependent failure models.
\newblock {\em Planet. Space. Sci.}, {\bf 107}, 3--9.

\bibitem[\protect\citename{Jutzi \& Asphaug, }2015]{Jutzi-and-Asphaug2015}
Jutzi, M., \& Asphaug, E. 2015.
\newblock The shape and structure of cometary nuclei as a result of
  low-velocity accretion.
\newblock {\em Science}, {\bf 348}, 1355--1358.

\bibitem[\protect\citename{{Knight} {\em et~al.\ }\relax,
  }2017]{Knight-et-al2017}
{Knight}, M.~M., {Protopapa}, S., {Kelley}, M.~S.~P., {Farnham}, T.~L.,
  {Bauer}, J.~M., {Bodewits}, D., {Feaga}, L.~M., \& {Sunshine}, J.~M. 2017.
\newblock On the Rotation Period and Shape of the Hyperbolic Asteroid
  {1I/‘Oumuamua} (2017 {U1}) from Its Lightcurve.
\newblock {\em Astrophys. J.}, {\bf 851}(2), L31.

\bibitem[\protect\citename{{Kobayashi}, }2015]{Kobayashi2015}
{Kobayashi}, H. 2015.
\newblock {Orbital evolution of planetesimals in gaseous disks}.
\newblock {\em Earth, Planets, and Space}, {\bf 67}(Apr.), 60.

\bibitem[\protect\citename{{Kobayashi} \& {Ida}, }2001]{Kobayashi-and-Ida2001}
{Kobayashi}, H., \& {Ida}, S. 2001.
\newblock {The Effects of a Stellar Encounter on a Planetesimal Disk}.
\newblock {\em Icarus}, {\bf 153}(Oct.), 416--429.

\bibitem[\protect\citename{{Kobayashi} {\em et~al.\ }\relax,
  }2005]{Kobayashi-et-al2005}
{Kobayashi}, H., {Ida}, S., \& {Tanaka}, H. 2005.
\newblock {The evidence of an early stellar encounter in Edgeworth Kuiper
  belt}.
\newblock {\em Icarus}, {\bf 177}(Sept.), 246--255.

\bibitem[\protect\citename{{Kretke} {\em et~al.\ }\relax,
  }2009]{Kretke-et-al2009}
{Kretke}, K.~A., {Lin}, D.~N.~C., {Garaud}, P., \& {Turner}, N.~J. 2009.
\newblock {Assembling the Building Blocks of Giant Planets Around
  Intermediate-Mass Stars}.
\newblock {\em Astrophys. J.}, {\bf 690}(Jan.), 407--415.

\bibitem[\protect\citename{{Leinhardt} {\em et~al.\ }\relax,
  }2000]{Leinhardt-et-al2000}
{Leinhardt}, Z.~M., {Richardson}, D.~C., \& {Quinn}, T. 2000.
\newblock {Direct N-body Simulations of Rubble Pile Collisions}.
\newblock {\em Icarus}, {\bf 146}(July), 133--151.

\bibitem[\protect\citename{{Leinhardt} {\em et~al.\ }\relax,
  }2010]{Leinhardt-et-al2010}
{Leinhardt}, Z.~M., {Marcus}, R.~A., \& {Stewart}, S.~T. 2010.
\newblock {The Formation of the Collisional Family Around the Dwarf Planet
  Haumea}.
\newblock {\em Astrophys. J.}, {\bf 714}(May), 1789--1799.

\bibitem[\protect\citename{{Libersky} \& {Petschek},
  }1991]{Libersky-and-Petschek1991}
{Libersky}, L.~D., \& {Petschek}, A.~G. 1991.
\newblock {Smooth particle hydrodynamics with strength of materials}.
\newblock {\em Pages  248--257 of:} {Trease}, H.~E., {Fritts}, M.~F., \&
  {Crowley}, W.~P. (eds), {\em Advances in the Free-Lagrange Method Including
  Contributions on Adaptive Gridding and the Smooth Particle Hydrodynamics
  Method}.
\newblock Lecture Notes in Physics, Berlin Springer Verlag, vol. 395.

\bibitem[\protect\citename{{Meech} {\em et~al.\ }\relax,
  }2017]{Meech-et-al2017}
{Meech}, K.~J., {Weryk}, R., {Micheli}, M., {Kleyna}, J.~T., {Hainaut}, O.~R.,
  {Jedicke}, R., {Wainscoat}, R.~J., {Chambers}, K.~C., {Keane}, J.~V.,
  {Petric}, A., {Denneau}, L., {Magnier}, E., {Berger}, T., {Huber}, M.~E.,
  {Flewelling}, H., {Waters}, C., {Schunova-Lilly}, E., \& {Chastel}, S. 2017.
\newblock {A brief visit from a red and extremely elongated interstellar
  asteroid}.
\newblock {\em Nature}, {\bf 552}(Dec.), 378--381.

\bibitem[\protect\citename{{Micheli} {\em et~al.\ }\relax,
  }2018]{Micheli-et-al2018}
{Micheli}, M., {Farnocchia}, D., {Meech}, K.~J., {Buie}, M.~W., {Hainaut},
  O.~R., {Prialnik}, D., {Sch{\"o}rghofer}, N., {Weaver}, H.~A., {Chodas},
  P.~W., {Kleyna}, J.~T., {Weryk}, R., {Wainscoat}, R.~J., {Ebeling}, H.,
  {Keane}, J.~V., {Chambers}, K.~C., {Koschny}, D., \& {Petropoulos}, A.~E.
  2018.
\newblock {Non-gravitational acceleration in the trajectory of 1I/2017 U1
  ('Oumuamua)}.
\newblock {\em Nature}, {\bf 559}(June), 223--226.

\bibitem[\protect\citename{{Michikami} {\em et~al.\ }\relax,
  }2016]{Michikami-et-al2016}
{Michikami}, T., {Hagermann}, A., {Kadokawa}, T., {Yoshida}, A., {Shimada}, A.,
  {Hasegawa}, S., \& {Tsuchiyama}, A. 2016.
\newblock Fragment shapes in impact experiments ranging from cratering to
  catastrophic disruption.
\newblock {\em Icarus}, {\bf 264}(Supplement C), 316 -- 330.

\bibitem[\protect\citename{{Monaghan} \& {Lattanzio},
  }1985]{Monaghan-and-Lattanzio1985}
{Monaghan}, J.~J., \& {Lattanzio}, J.~C. 1985.
\newblock {A refined particle method for astrophysical problems}.
\newblock {\em Astron. Astrophys.}, {\bf 149}(Aug.), 135--143.

\bibitem[\protect\citename{{Mori} {\em et~al.\ }\relax, }2017]{Mori-et-al2017}
{Mori}, S., {Muranushi}, T., {Okuzumi}, S., \& {Inutsuka}, S.-i. 2017.
\newblock {Electron Heating and Saturation of Self-regulating Magnetorotational
  Instability in Protoplanetary Disks}.
\newblock {\em Astrophys. J.}, {\bf 849}(Nov.), 86.

\bibitem[\protect\citename{{Ohtsuki} {\em et~al.\ }\relax,
  }2002]{Ohtsuki-et-al2002}
{Ohtsuki}, K., {Stewart}, G.~R., \& {Ida}, S. 2002.
\newblock {Evolution of Planetesimal Velocities Based on Three-Body Orbital
  Integrations and Growth of Protoplanets}.
\newblock {\em Icarus}, {\bf 155}(Feb.), 436--453.

\bibitem[\protect\citename{{Ormel} \& {Cuzzi}, }2007]{Ormel-and-Cuzzi2007}
{Ormel}, C.~W., \& {Cuzzi}, J.~N. 2007.
\newblock {Closed-form expressions for particle relative velocities induced by
  turbulence}.
\newblock {\em Astron. Astrophys.}, {\bf 466}(May), 413--420.

\bibitem[\protect\citename{{Pinte} {\em et~al.\ }\relax,
  }2016]{Pinte-et-al2016}
{Pinte}, C., {Dent}, W.~R.~F., {M{\'e}nard}, F., {Hales}, A., {Hill}, T.,
  {Cortes}, P., \& {de Gregorio-Monsalvo}, I. 2016.
\newblock {Dust and Gas in the Disk of {HL Tauri}: Surface Density, Dust
  Settling, and Dust-to-gas Ratio}.
\newblock {\em Astrophys. J.}, {\bf 816}(Jan.), 25.

\bibitem[\protect\citename{{Raymond} {\em et~al.\ }\relax,
  }2018]{Raymond-et-al2017}
{Raymond}, S.~N., {Armitage}, P.~J., {Veras}, D., {Quintana}, E.~V., \&
  {Barclay}, T. 2018.
\newblock {Implications of the interstellar object {1I/'Oumuamua} for planetary
  dynamics and planetesimal formation}.
\newblock {\em Mon. Not. R. Astron. Soc.}, {\bf 476}(May), 3031--3038.

\bibitem[\protect\citename{{Richardson} {\em et~al.\ }\relax,
  }2002]{Richardson-et-al2002}
{Richardson}, D.~C., {Leinhardt}, Z.~M., {Melosh}, H.~J., {Bottke}, Jr., W.~F.,
  \& {Asphaug}, E. 2002.
\newblock {\em {Gravitational Aggregates: Evidence and Evolution}}.
\newblock Pages  501--515.

\bibitem[\protect\citename{{Richardson} {\em et~al.\ }\relax,
  }2005]{Richardson-et-al2005}
{Richardson}, D.~C., {Elankumaran}, P., \& {Sanderson}, R.~E. 2005.
\newblock {Numerical experiments with rubble piles: equilibrium shapes and
  spins}.
\newblock {\em Icarus}, {\bf 173}(Feb.), 349--361.

\bibitem[\protect\citename{Sugiura {\em et~al.\ }\relax,
  }2018]{Sugiura-et-al2018}
Sugiura, K., Kobayashi, H., \& Inutsuka, S. 2018.
\newblock {Toward understanding the origin of asteroid geometries. {V}ariety in
  shapes produced by equal-mass impacts}.
\newblock {\em Astron. Astrophys.}, {\bf 620}, A167.

\bibitem[\protect\citename{{Takahashi} \& {Inutsuka},
  }2014]{Takahashi-and-Inutsuka2014}
{Takahashi}, S.~Z., \& {Inutsuka}, S.-i. 2014.
\newblock {Two-component Secular Gravitational Instability in a Protoplanetary
  Disk: A Possible Mechanism for Creating Ring-like Structures}.
\newblock {\em Astrophys. J.}, {\bf 794}(Oct.), 55.

\bibitem[\protect\citename{{Takahashi} \& {Inutsuka},
  }2016]{Takahashi-and-Inutsuka2016}
{Takahashi}, S.~Z., \& {Inutsuka}, S.-i. 2016.
\newblock {An Origin of Multiple Ring Structure and Hidden Planets in HL Tau: A
  Unified Picture by Secular Gravitational Instability}.
\newblock {\em Astron. J.}, {\bf 152}(Dec.), 184.

\bibitem[\protect\citename{{Tominaga} {\em et~al.\ }\relax,
  }2018]{Tominaga-et-al2018}
{Tominaga}, R.~T., {Inutsuka}, S.-i., \& {Takahashi}, S.~Z. 2018.
\newblock {Non-linear development of secular gravitational instability in
  protoplanetary disks}.
\newblock {\em Publ. Astron. Soc. Jpn.}, {\bf 70}(Jan.), 3.

\bibitem[\protect\citename{{{\v D}urech} {\em et~al.\ }\relax,
  }2012]{Durech-et-al2012}
{{\v D}urech}, J., {Vokrouhlick{\'y}}, D., {Baransky}, A.~R., {Breiter}, S.,
  {Burkhonov}, O.~A., {Cooney}, W., {Fuller}, V., {Gaftonyuk}, N.~M., {Gross},
  J., {Inasaridze}, R.~Y., {Kaasalainen}, M., {Krugly}, Y.~N., {Kvaratshelia},
  O.~I., {Litvinenko}, E.~A., {Macomber}, B., {Marchis}, F., {Molotov}, I.~E.,
  {Oey}, J., {Polishook}, D., {Pollock}, J., {Pravec}, P., {S{\'a}rneczky}, K.,
  {Shevchenko}, V.~G., {Slyusarev}, I., {Stephens}, R., {Szab{\'o}}, G.,
  {Terrell}, D., {Vachier}, F., {Vanderplate}, Z., {Viikinkoski}, M., \&
  {Warner}, B.~D. 2012.
\newblock {Analysis of the rotation period of asteroids (1865) {C}erberus,
  (2100) {R}a-{S}halom, and (3103) {E}ger - search for the {YORP} effect}.
\newblock {\em Astron. Astrophys.}, {\bf 547}(Nov.), A10.

\bibitem[\protect\citename{{Wakita} {\em et~al.\ }\relax,
  }2014]{Wakita-et-al2014}
{Wakita}, S., {Nakamura}, T., {Ikeda}, T., \& {Yurimoto}, H. 2014.
\newblock {Thermal modeling for a parent body of Itokawa}.
\newblock {\em Meteoritics and Planetary Science}, {\bf 49}(Feb.), 228--236.

\bibitem[\protect\citename{Walsh \& Richardson,
  }2006]{Walsh-and-Richardson2006}
Walsh, K.~J., \& Richardson, D.~C. 2006.
\newblock Binary near-Earth asteroid formation: Rubble pile model of tidal
  disruptions.
\newblock {\em Icarus}, {\bf 180}(1), 201 -- 216.

\bibitem[\protect\citename{{Weidenschilling}, }1977]{Weidenschilling1977}
{Weidenschilling}, S.~J. 1977.
\newblock {Aerodynamics of solid bodies in the solar nebula}.
\newblock {\em MNRAS}, {\bf 180}(July), 57--70.

\bibitem[\protect\citename{{Ye} {\em et~al.\ }\relax, }2017]{Ye-et-al2017}
{Ye}, Q.-Z., Zhang, Q., Kelley, M. S.~P., \& Brown, P.~G. 2017.
\newblock {1I/2017 U1 (‘Oumuamua)} is Hot: {I}maging, Spectroscopy, and
  Search of Meteor Activity.
\newblock {\em Astrophys. J.}, {\bf 851}(1), L5.

\end{thebibliography}

\section*{Appendix A. Validity of reduced sound speed}

Here, we show the validity of artificially reduced sound speed. Motion of each SPH particle is determined by the self-gravity, pressure distribution, and deviatoric stress distribution. If impact velocities are much smaller than the sound speed, pressure distribution is determined by the self-gravity because pressure induced by the impact is immediately relaxed. In this study deviatoric stress is determined by the friction, and frictional force is proportional to the pressure; deviatoric stress distribution is also determined by the self-gravity. Therefore, motion of each SPH particle is solely determined by the self-gravity, and thus the sound speed does not affect results of impacts.

\begin{figure}[!htb]
  \begin{center}
    \includegraphics[bb=0 0 974 374, width=1.0\linewidth,clip]{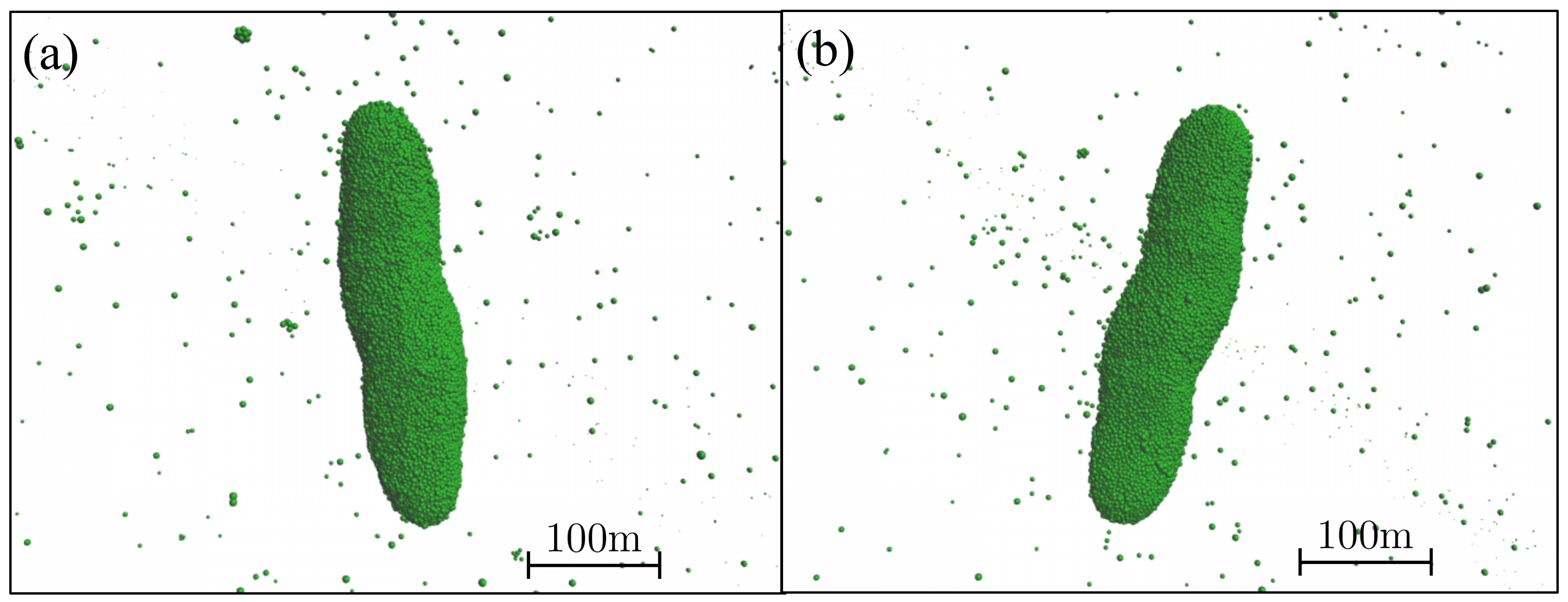}
    \caption{Snapshots of the impact with $R_{{\rm t}}=50\,{\rm m}$, $q=1$, $\phi_{{\rm d}}=40^{\circ}$, $\theta_{{\rm imp}}=15^{\circ}$, and $v_{{\rm imp}}=20\,{\rm cm/s}$ at $t=1.0\times 10^{5}\,{\rm s}$. (a) shows the result of the simulation with $A_{{\rm Til}}=B_{{\rm Til}}=2.67\times 10^{4}\,{\rm Pa}$, and (b) shows that with $A_{{\rm Til}}=B_{{\rm Til}}=2.67\times 10^{5}\,{\rm Pa}$.}
    \label{50m-1.0-15-20-differentModulus}
  \end{center}
\end{figure}

We conduct simulations of the impact with $R_{{\rm t}}=50\,{\rm m}$, $q=1$, $\theta_{{\rm imp}}=15^{\circ}$, $\phi_{{\rm d}}=40^{\circ}$, $v_{{\rm imp}}=20\,{\rm cm/s}$, and two different values of sound speed, or Tillotson parameters of $A_{{\rm Til}}$ and $B_{{\rm Til}}$. Figs.\,\ref{50m-1.0-15-20-differentModulus}a and b show snapshots of the impact outcomes with $A_{{\rm Til}}=B_{{\rm Til}}=2.67\times 10^{4}\,{\rm Pa}$ and $A_{{\rm Til}}=B_{{\rm Til}}=2.67\times 10^{5}\,{\rm Pa}$, respectively. From Fig.\,\ref{50m-1.0-15-20-differentModulus}, we find that results of simulations become almost the same even if values of $A_{{\rm Til}}$ and $B_{{\rm Til}}$ are ten times different.

\end{document}